\documentclass[fleqn,usenatbib]{mnras}

\usepackage[T1]{fontenc}
\usepackage{ae,aecompl}

\usepackage{graphicx}	
\usepackage{amsmath}	
\usepackage{amssymb}	
\usepackage{times,txfonts} 
\usepackage[normalem]{ulem}
\usepackage{lineno}

\title[Intracluster Light in XCS-HSC]{The Growth of Intracluster Light in XCS-HSC Galaxy Clusters from $0.1 < z < 0.5$}

\author[K. E. Furnell et al.]{Kate E. Furnell$^{1}$\thanks{E-mail: k.e.furnell@liverpool.ac.uk},
Chris A. Collins$^{1}$,
Lee S. Kelvin$^{1,2}$,
Ivan K. Baldry$^{1}$,
Phil A. James$^{1}$, \newauthor
Maria Manolopoulou$^{3}$, 
Robert G. Mann$^{3}$, 
Paul A. Giles$^{4}$, 
Alberto  Bermeo$^{4}$, \newauthor
Matthew Hilton$^{5,6}$, 
Reese Wilkinson$^{4}$,
A. Kathy Romer$^{4}$,
Carlos Vergara$^{4}$, \newauthor
Sunayana Bhargava$^{4}$, 
John P. Stott$^{7}$,
Julian Mayers$^{4}$, 
and Pedro Viana$^{8,9}$. 
\\
$^{1}$Astrophysics Research Institute, Liverpool John Moores University, IC2, Liverpool Science Park, 146 Brownlow Hill, Liverpool, L3 5RF\\
$^{2}$Department of Astrophysical Sciences, Princeton University, 4 Ivy Ln, Princeton, NJ 08544, United States, USA\\
$^{3}$Institute for Astronomy, University of Edinburgh, Blackford Hill, EH9 3HJ, Edinburgh, UK\\
$^{4}$Department of Physics and Astronomy, University of Sussex, Brighton, BN1 9QH, UK\\
$^{5}$Astrophysics Research Centre, University of KwaZulu-Natal, Westville Campus, Durban 4041, South Africa \\
$^{6}$School of Mathematics, Statistics \& Computer Science, University of KwaZulu-Natal, Westville Campus, Durban 4041, South Africa \\
$^{7}$Department of Physics, Lancaster University, Lancaster, LA1 4YB, UK \\
$^{8}$Instituto de Astrof\'{\i}sica e Ci\^{e}ncias do Espa\c{c}o, Universidade do Porto, CAUP, Rua das Estrelas, 4150-762 Porto, Portugal\\
$^{9}$Departamento de F\'{\i}sica e Astronomia, Faculdade de Ci\^{e}ncias, Universidade do Porto, Rua do Campo Alegre, 687, 4169-007 Porto, Portugal \\
}

\date{Accepted 05 January 2021; received 04 January 2021; in original form 04 December 2019}

\pubyear{2021}

\begin{document}
\label{firstpage}
\pagerange{\pageref{firstpage}--\pageref{lastpage}}
\maketitle
\begin{abstract}
We estimate the Intracluster Light (ICL) component within a sample of 18 clusters detected in XMM Cluster Survey (XCS) data using deep ($\sim$ 26.8 mag) Hyper Suprime Cam Subaru Strategic Program DR1 (HSC-SSP DR1) $i$-band data. We apply a rest-frame ${\mu}_{B} = 25 \ \mathrm{mag/arcsec^{2}}$ isophotal threshold to our clusters, below which we define light as the ICL within an aperture of $R_{X,500}$ (X-ray estimate of $R_{500}$) centered on the Brightest Cluster Galaxy (BCG). After applying careful masking and corrections for flux losses from background subtraction, we recover $\sim$20\% of the ICL flux, approximately four times our estimate of the typical background at the same isophotal level ($\sim 5\%$). We find that the ICL makes up about $\sim 24\%$ of the total cluster stellar mass on average ($\sim$ 41\% including the flux contained in the BCG within 50 kpc); this value is well-matched with other observational studies and semi-analytic/numerical simulations, but is significantly smaller than results from recent hydrodynamical simulations (even when measured in an observationally consistent way). We find no evidence for any links between the amount of ICL flux with cluster mass, but find a growth rate of $2-4$ for the ICL between $0.1 < z < 0.5$. We conclude that the ICL is the dominant evolutionary component of stellar mass in clusters from $z \sim 1$. Our work highlights the need for a consistent approach when measuring ICL alongside the need for deeper imaging, in order to unambiguously measure the ICL across as broad a redshift range as possible (e.g. 10-year stacked imaging from the Vera C. Rubin Observatory).
\end{abstract}




\section{Introduction}
\label{intro}
\par A complete understanding of the growth of universal large-scale structure (LSS) is one of the primary goals of modern cosmology. Structures which make up the `cosmic web' include: `nodes' (gravitationally-bound groups and clusters of galaxies), `filaments' (lower-density connective `strings' of galaxies) and `voids' (vast under-densities of galaxies). These have been observed extensively in nature, initially by Fritz Zwicky, with widespread cataloguing later by individuals such as George O. Abell in the early-to-mid 20th century (e.g. \citealt{1937ApJ....86..217Z}; \citealt{1958ApJS....3..211A}) to a more extensive scale by modern spectroscopic galaxy surveys (e.g. 2dFGRS, \citealt{2001MNRAS.328.1039C}). Our comprehension of how matter - baryonic (protons, neutrons, electrons) and dark - collapses to form these structures (and the rate at which this happens) is partially governed by our understanding of cosmology (e.g. BAHAMAS, \citealt{2018MNRAS.476.2999M}).
\par Effective comparisons between observed cluster properties and outputs from hydrodynamical simulations remain critical when attempting to accurately model LSS. In recent years, cosmological hydrodynamical simulations have been reasonably successful in reproducing the structures observed in nature (e.g. Millennium, \citealt{2005Natur.435..629S}, see their Figure 1). However, for example, at individual cluster scales, there are numerous key inconsistencies (e.g. the baryonic matter fraction). This has motivated higher-resolution `zoom' simulations with more complex `subgrid' physics to better understand these differences (e.g. \citealt{2017MNRAS.471.1088B}), as well as applying semi-analytic models (SAMs) to simulated dark matter haloes (e.g. \citealt{2007MNRAS.375....2D}). 
\par These discrepancies are especially striking in the case of Brightest Cluster Galaxies (BCGs); massive, often non-star forming galaxies which primarily reside at the X-ray peak of galaxy clusters, a proxy used for the bottom of the gravitational potential well (e.g. \citealt{2004ApJ...617..879L}). For example, there are unresolved tensions with most cosmological simulations regarding `profile cuspiness' (e.g. \citealt{1996ApJ...462..563N}), with observed BCGs having a `core' present in their dark-matter density profiles (e.g. \citealt{2013ApJ...765...25N}) which cannot readily exist in the $\Lambda$CDM paradigm for non-self-interacting dark matter (e.g. \citealt{2017MNRAS.472.1972H}). Vitally, there are also tensions present between the observed stellar mass growth rate of BCGs (e.g. \citealt{2009Natur.458..603C}; \citealt{2012MNRAS.425.2058B}) and that in simulations (e.g. \citealt{2007MNRAS.375....2D}; \citealt{2013MNRAS.435..901L}), with simulations generally predicting significantly more rapid rates of growth ($\sim 2-4 \times$ since $z \sim 1$) than those observed in nature (although significant improvements with better agreement have been made in recent studies, e.g. \citealt{2018MNRAS.479.1125R}).
\par One of the proposed `solutions' to this missing BCG stellar mass problem is analysis of the co-evolution of cluster BCGs with the Intracluster Light (ICL, e.g. \citealt{1952PASP...64..242Z}; \citealt{1972ApJ...176....1G}; \citealt{2011ApJS..195...15D} and numerous others) which is a low-surface brightness (LSB; <1\% sky level, e.g. \citealt{2017MNRAS.468.2569B}), diffuse stellar component in clusters. The origin of the ICL is debated extensively in the literature, namely whether it originates primarily from BCG-passive satellite mergers (e.g. \citealt{2013MNRAS.434.2856B}; \citealt{2005ApJ...618..195G}), tidal stripping from infalling, younger satellites (e.g. \citealt{2018MNRAS.474.3009D}; \citealt{2017ApJ...846..139M} \citealt{2015MNRAS.448.1162D}; \citealt{2014ApJ...794..137M}; \citealt{2018MNRAS.474..917M}), in-situ star formation due to ICM collapse in the case of gas-rich clusters (\citealt{2010MNRAS.406..936P}), or a combination of these.
\par Exactly how much the ICL contributes to the stellar mass of a cluster at a given epoch is much debated throughout the literature. At present epochs ($z \sim 0$), observational results span over a wide range of values (10-50\%) with the same being true for simulations; significant tension also however exists between them with respect to the rate of observed ICL growth (e.g. \citealt{2007MNRAS.377....2M}, \citealt{2010MNRAS.405.1544D}, \citealt{2011ApJ...732...48R}, \citealt{2014MNRAS.437.3787C} and \citealt{2018ApJ...859...85T}). The reasons behind these deviations are unclear, with sample selection, data quality and method of measurement all being contributing factors to the scatter. As the ICL is a faint component which is not bound to any one cluster galaxy, a concise definition in an observational context is non-trivial. 
Some authors attempt to model the light profiles of galaxies to disentangle their haloes from the true ICL (e.g. \citealt{2017ApJ...846..139M}; \citealt{2007ApJ...666..147G}), whereas others use a isophotal thresholding technique (e.g. \citealt{2012MNRAS.425.2058B}; \citealt{2015MNRAS.449.2353B}) or use ellipsoidal masks derived from basic structural parameters to mask cluster objects (e.g. \citealt{1980ApJS...43..305K}; see \citealt{2005MNRAS.358..949Z}, \citealt{2018MNRAS.474.3009D}); or use a wavelet-like approach (e.g. \citealt{2005MNRAS.364.1069D, 2008MNRAS.388.1433D};  \citealt{2015IAUGA..2253756J, 2018ApJ...857...79J}; \citealt{2019A&A...628A..34E}). All of these methods have various biases and caveats.
\par In this work, we study the ICL component of a sample of X-ray selected galaxy clusters from the XMM-Cluster Survey, using deep ($i \sim 26.8$ mag, or 28.3 mag/arcsec$^{2}$ (5$\sigma$, $2^{\prime\prime} \times 2^{\prime\prime}$)) Hyper Suprime-Cam Strategic Survey Program DR1 imaging (\citealt{2018PASJ...70S...8A}). In doing so, we hope to gain a greater understanding of the nature of the accumulation of stellar mass in the cores of clusters since $z \sim 0.5$.
This paper is structured as follows: firstly, we discuss the parent sample of the clusters used for this study; secondly, we outline our selection and detail our methodology used in quantifying the ICL; lastly, we discuss our results. We adopt, where applicable, a standard  $\mathrm{\Lambda}$CDM concordance cosmology throughout, with $\mathrm{H_{0}} = 70 \ \mathrm{km \ s^{-1}Mpc^{-1}}$, $\mathrm{h_{100} \ = \ 0.7}$, $\mathrm{{\Omega}_{\Lambda}} = 0.7$ and $\mathrm{{\Omega}_{M}} = 0.3$.
\section{Data}
\subsection{The XMM Cluster Survey}
The XMM Cluster Survey (\citealt{2001ApJ...547..594R}) is an all-sky serendipitous search for galaxy clusters using legacy X-ray data from the XMM-Newton space telescope (e.g. \citealt{2001A&A...365L...1J}). The first XCS data release in 2012 (\citealt{2012MNRAS.423.1024M}) contained X-ray and optical confirmations for 503 galaxy clusters, a third of which were entirely new to the literature. The second XCS public data release (Giles et al., in prep) increases the number of clusters detected in XCS to $\sim 1300$ and overlap with this master catalogue in HSC forms the basis of the sample we use in this work\footnote{A comparison between the HSC footprint and other surveys can be found here: \url{https://hsc.mtk.nao.ac.jp/ssp/survey/}.}. Due to the considerably less biased means of cluster selection in X-rays than optical surveys coupled with high angular resolution X-ray imaging ($4.1^{\prime\prime}$), XCS data is ideal for constructing a representative cluster sample.
\par In the case of the sample used in this work (Giles et al., in prep), XCS detections were cross-matched for spectroscopy with the SDSS DR13, VIPERS PDR2 and DEEP2 surveys (\citealt{2017ApJS..233...25A}; \citealt{2014A&A...566A.108G} and \citealt{2013ApJS..208....5N} respectively). Spectroscopic redshifts are assigned to each cluster through application of a biweight location estimator (see \citealt{1990AJ....100...32B}) using all galaxies falling within $1.5^{\prime}$ from the XCS centroid from the X-ray Automated Pipeline Algorithm (\textsc{XAPA}, \citealt{2011MNRAS.418...14L}); this redshift centroid is then re-calculated after applying a clip of $\Delta v \pm 3000 \ \mathrm{kms^{-1}}$ about the initial redshift, within a radius of 1.5 Mpc projected distance from the XAPA centroid (see method described in \citealt{2018ApJS..235...20H}). Section \ref{sec:sampsel} details the outcome of the matching process for the sample used here.
\section{Hyper Suprime-Cam Subaru Strategic Program}\label{hsc}
\subsection{Survey Description}
In this work, we make use of optical imaging data from the first release of the Hyper Suprime-Cam Subaru Strategic Program (HSC-SSP, e.g. \citealt{2018PASJ...70S...4A}); one of the deepest, public ground-based optical surveys available (see Table \ref{tab:HSC_brightlim}). The HSC instrument is a wide-field (1.8 \ $\mathrm{deg^{2}}$) imaging camera on the 8.2 m Subaru telescope on Mauna Kea, Hawaii where the SSP has been running since March, 2014. In total, the SSP is scheduled for a run of 300 nights over the course of six years, covering three imaging depths in total: `Wide', `Deep' and `Ultra-Deep' in 5 Sloan-like passbands ($grizy$). In this work, we use imaging from the `Deep' subset, chosen to keep the imaging data for our cluster sample as consistently deep as possible. A summary table of the average $5 \sigma$ limiting depths has been included for reference for the available runs and broad bands (Table \ref{tab:HSC_brightlim}). The survey footprint overlaps with numerous other surveys, such as the general Sloan footprint and its associated surveys (e.g. \citealt{2000AJ....120.1579Y}), Pan-STARRS (e.g. \citealt{2016arXiv161205560C}), COSMOS (e.g. \citealt{2007ApJS..172....1S}) and DEEP-2 (e.g. \citealt{2013ApJS..208....5N}). The imaging depth of HSC far exceeds that of any current public survey, with the exception of the Hubble Frontier Fields (HFF, \citealt{2017ApJ...837...97L}) e.g. KiDS (\citealt{2013ExA....35...25D}), DES, (\citealt{2005astro.ph.10346T}). Current estimates of HSC image quality are comparable to surveys anticipated by the upcoming Vera C. Rubin Observatory (formerly the Large Synoptic Survey Telescope, see \citealt{2008arXiv0805.2366I} and \citealt{brough2020vera}, also Section \ref{lsst:pipe} for further comments on data reduction).
\subsection{Data Reduction}\label{lsst:pipe}
\begin{table}
\centering
	\caption{A summary table of the average limiting depths for the HSC-SSP survey. In this work, we use the `Deep' layer in the $i$-band (DR1 area $\sim \mathrm{26} \ \mathrm{deg^{2}}$).}
\begin{tabular}{l|l|l}
Layer   \hfill   & Filter \hfill  & Lim. mag. ($5 \sigma$, $2^{\prime\prime}$) \hfill \\ \hline
Wide   \hfill    & $g$, $r$ \hfill & 26.5, 26.1        \hfill          \\
Wide    \hfill   & $i$    \hfill  & 25.9                \hfill        \\
Wide    \hfill   & $z$, $y$ \hfill & 25.1, 24.4         \hfill         \\
Deep    \hfill   & $g$, $r$ \hfill & 27.5, 27.1         \hfill         \\
Deep   \hfill    & $i$     \hfill & 26.8                \hfill        \\
Deep   \hfill    & $z$     \hfill & 26.3                \hfill        \\
Deep     \hfill  & $y$     \hfill & 25.3                \hfill        \\
Ultra Deep \hfill & $g$, $r$ \hfill & 28.1, 27.7        \hfill          \\
Ultra Deep \hfill & $i$     \hfill & 27.4               \hfill         \\
Ultra Deep \hfill & $z$, $y$ \hfill & 26.8, 26.3         \hfill        
\end{tabular}
\label{tab:HSC_brightlim}
\end{table}
For the DR1 release, the HSC-SSP data products have undergone processing through the HSC pipeline, an adapted version of the Vera C. Rubin Observatory Data Management (DM) software stack in preparation for Vera C. Rubin Observatory data products in the coming decade (see \citealt{2015arXiv151207914J} for a description of the Vera C. Rubin Observatory DM stack). The full implementation for HSC is detailed in \cite{2018PASJ...70S...5B} (including a flow diagram of the complete process, see their Figure 1) but we include an abridged version here to provide context. The pipeline software itself is open source and licensed for public use under the GNU public license (version 3). The photometric performance of the pipeline on mock objects is described in detail in \cite{2018PASJ...70S...6H}, who demonstrate a strong recovery in input versus output flux even for de Vaucouleurs-like objects (on average $\sim 85 \%$ at $m_{i} = 25$). They acknowledge, however, that the HSC pipeline tends to over-subtract flux around extended, bright objects (which they explore further when studying the faint haloes of elliptical-type galaxies in \citealt{2018MNRAS.475.3348H}). We discuss this issue, along with a proposition of a post-processing `fix', in Section \ref{section:divcorr}.
\par In simplified terms, much of the HSC pipeline is built on algorithms and concepts originating from the SDSS $photo$ pipeline (see \citealt{2001ASPC..238..269L}), the pipeline which produces the data products for all SDSS data releases. Raw data and coadds can be queried online on the HSC-SSP DR1 release site; alternatively, there are reduced data products (e.g. photometry, best-fit models, photo-$z$ estimates) available which can be downloaded via SQL query.
\par The HSC pipeline operates in several stages to produce the final scientific data products. The process (with relevant details) is roughly as follows: 
\begin{enumerate}
\item \textbf{CCD processing:} the raw data from each CCD is taken, and basic data corrections and calibrations are applied. Firstly, an Instrument Signature Removal (ISR) is applied, which embodies basic reduction (i.e. flat, bias and dark corrections), brighter-fatter corrections (for source intensity dependence on the measured PSF), corrections for crosstalk and corrections for CCD non-linearity (e.g. \citealt{2007AJ....134..466K} for context as to how this applies to ICL). The sky is estimated for each image and subtracted using a variance-weighted $6^{\mathrm{th}}$ order Chebyshev polynomial sampled over $128 \times 128$ 3$\sigma$ clipped average pixel values. 
\par In summary, this stage produces two main data products: calibrated exposure data (i.e. datacubes which contain: a background subtracted, calibrated image; a mask frame containing source detections, pixel flags and star masks; a variance frame, essentially a `weight map' describing the pixel-by pixel variance of the coadded images) and a `source catalogue', namely a database of detected objects with photometric information as measured by the pipeline.
\item \textbf{Joint calibration:} when all CCDs have been processed, their astrometric and photometric calibrations are refined by requiring consistent positional and flux values of sources on repeat visits where they may appear on different regions of the focal plane.
\item \textbf{Image coaddition:} the individual CCD exposures are then coadded to improve the imaging depth. As is widely known in astronomical surveys, co-addition can lead to complications, such as data degradation or introduction of systematic errors. Efforts have been made during the HSC pipeline's construction to avoid these issues wherever possible; as stressed by \cite{2018PASJ...70S...5B}, the pipeline is still actively undergoing refinement.
\item \textbf{Coadd processing:} after creating the coadds, the pipeline carries out another round of image processing. Objects on the coadds are detected, deblended and measured, creating a catalogue of final object measurements. A final background is then subtracted for each sky `patch' via an average from a $4\mathrm{k} \times 4\mathrm{k}$ pixel bin.
\end{enumerate}
\section{Sample Selection}
\label{sec:sampsel}
\begin{figure}
\centering
	\includegraphics[width=\columnwidth]{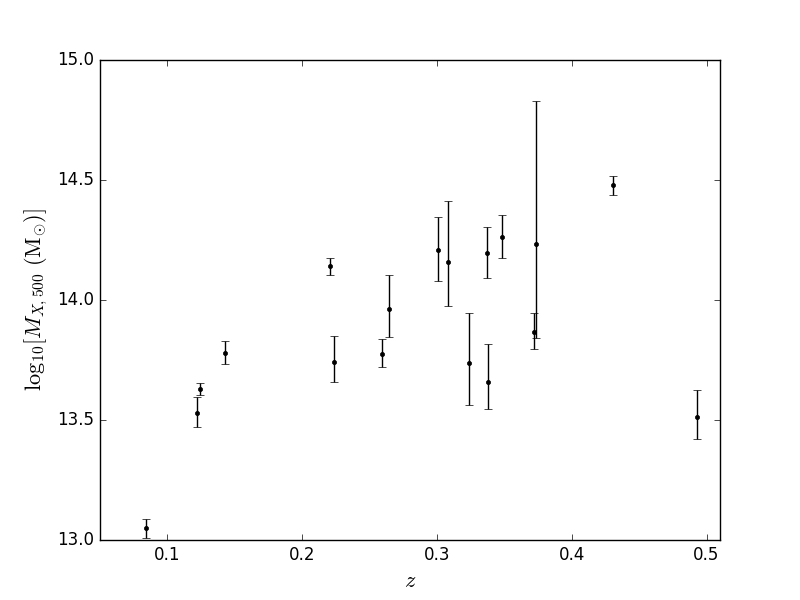}
    \caption{The $M_{X,500}-z$ relation for the clusters used in this work (see text for details). The redshifts are spectroscopic, with errors $\Delta z \sim 10^{-5}$. The clusters span a wide range in both redshift and mass; a correlation is detected, but it is not significant (see table \ref{tab:proj2fullspear}).}
    \label{fig:lx_z}
\end{figure}
To create our sample of clusters, we cross-matched the XCS-DR2 North (Giles et al., in prep) master source list with the entire HSC-SSP DR1 footprint region (Wide, Deep and Ultra-Deep). This produced an initial match of 202 common sources. We required, for robustness, for there to be an available spectroscopic redshift for both the assigned BCG and for the cluster itself; 79 objects met this criterion. The BCGs in this work are assigned through the GMPhoRCC algorithm of \cite{2017MNRAS.469.3851H} and then eyeballed individually using optical images with overlaid X-ray contours. The GMPhorCC algorithm models galaxy distributions as Gaussian mixtures using SDSS DR10 data, using objects from the main galaxy catalogue (see paper for details on colour selection criteria and identifying the red sequence; see also their Figure 4 for a detailed flowchart on the operation of the algorithm). From this, it was decided that no reassignments were necessary. 
\par The BCG and cluster spectroscopic redshifts were then compared - if they deviated significantly from one another beyond a specified velocity space limit ($\Delta v > \pm 5000 \ \mathrm{kms^-1}$), these objects were discarded (8 objects, leaving 71). We then required that each cluster had X-ray source parameter measurements (e.g. X-ray temperature, $T_{X,500}$) from XAPA (53 objects). Finally, to ensure the depth of our images were approximately consistent, we selected only sources which lay within the HSC-SSP Deep footprint (29 objects). 
\begin{table*}
\centering
\caption{The main parameters of the 18 XCS-HSC clusters used in this work. The BCG rest-frame $i$-band absolute magnitudes ($M_{i}$) are derived from aperture values as described in Section \ref{bcg:photo}. The relative errors are derived using the HSC variance maps and are typically quite small ($\Delta M_{\mathrm{i}} < 0.01$ mag).}
\begin{tabular}{l|l|l|l|l|l|l|l}
XCS ID                    & $\alpha_{2000}$ & $\delta_{2000}$ & $z$    & $M_{i}$ & $T_{X,500} \ (\mathrm{keV})$      & $R_{X,500}$ (Mpc)             & $M_{X,500} \ \mathrm{(10^{14} \times M_{\odot})}$ \\ \hline
XMMXCS J022456.1-050802.0 & 36.234          & -5.134          & 0.0840 & -23.023            & 0.648   ${\pm} 0.034$             & 0.331 ${\pm} 0.010$           & 0.112 ${\pm} 0.010$                               \\
XMMXCS J161039.2+540604.0 & 242.664         & +54.101         & 0.339  & -23.718            & 1.595   ${\pm}_{-0.227}^{+0.373}$ & 0.483 ${\pm}_{0.062}^{0.041}$ & 0.457 ${\pm}_{0.105}^{0.198}$                     \\
XMMXCS J233137.8+000735.0 & 352.908         & +0.126          & 0.224  & -23.690            & 1.719   ${\pm}_{-0.184}^{+0.269}$ & 0.537 ${\pm}_{0.046}^{0.033}$ & 0.553 ${\pm}_{0.971}^{0.156}$                     \\
XMMXCS J232923.6-004854.7 & 352.348         & -0.815          & 0.300  & -23.882            & 3.292   ${\pm}_{-0.524}^{+0.677}$ & 0.746 ${\pm}_{0.084}^{0.070}$ & 1.611 ${\pm}_{0.413}^{0.608}$                     \\
XMMXCS J161134.1+541640.5 & 242.892         & +54.278         & 0.337  & -24.009            & 3.278   ${\pm}_{-0.429}^{+0.511}$ & 0.729 ${\pm}_{0.063}^{0.056}$ & 1.567 ${\pm}_{0.334}^{0.441}$                     \\
XMMXCS J095902.7+025544.9 & 149.761         & +2.929          & 0.349  & -23.534            & 3.609   ${\pm}_{-0.400}^{+0.472}$ & 0.765 ${\pm}_{0.056}^{0.050}$ & 1.836 ${\pm}_{0.335}^{0.429}$                     \\
XMMXCS J095901.2+024740.4 & 149.755         & +2.794          & 0.501  & -23.587            & 1.385   ${\pm}_{-0.167}^{+0.223}$ & 0.406 ${\pm}_{0.036}^{0.029}$ & 0.327 ${\pm}_{0.064}^{0.095}$                     \\
XMMXCS J100141.6+022538.8 & 150.424         & +2.427          & 0.124  & -23.752            & 1.427   ${\pm}_{-0.045}^{+0.049}$ & 0.509 ${\pm} 0.010 $          & 0.424 ${\pm}_{0.022}^{0.025}$                     \\
XMMXCS J095737.1+023428.9 & 149.405         & +2.575          & 0.373  & -24.652            & 3.500   ${\pm}_{-1.443}^{+4.291}$ & 0.741 ${\pm}_{0.423}^{0.194}$ & 1.716 ${\pm}_{1.025}^{5.027}$                     \\
XMMXCS J022156.8-054521.9 & 35.487          & -5.756          & 0.259  & -23.619            & 1.814   ${\pm}_{-0.129}^{+0.157}$ & 0.544 ${\pm}_{0.026}^{0.022}$ & 0.595 ${\pm}_{0.071}^{0.091}$                     \\
XMMXCS J022148.1-034608.0 & 35.450          & -3.769          & 0.432  & -23.963            & 4.949   ${\pm}_{-0.245}^{+0.278}$ & 0.873 ${\pm}_{0.028}^{0.025}$ & 3.001 ${\pm}_{0.250}^{0.294}$                     \\
XMMXCS J022530.8-041421.1 & 36.378          & -4.239          & 0.143  & -23.294            & 1.761   ${\pm}_{-0.103}^{+0.122}$ & 0.568 ${\pm}_{0.022}^{0.019}$ & 0.602 ${\pm}_{0.059}^{0.073}$                     \\
XMMXCS J100047.3+013927.8 & 150.197         & +1.658          & 0.221  & -23.710            & 2.933   ${\pm}_{-0.137}^{+0.143}$ & 0.730 ${\pm}_{0.020}^{0.019}$ & 1.382 ${\pm}_{0.108}^{0.117}$                     \\
XMMXCS J022726.5-043207.1 & 36.861          & -4.535          & 0.308  & -23.662            & 3.090   ${\pm}_{-0.677}^{+1.273}$ & 0.716 ${\pm}_{0.160}^{0.100}$ & 1.438 ${\pm}_{0.496}^{1.156}$                     \\
XMMXCS J022524.8-044043.4 & 36.353          & -4.679          & 0.264  & -23.244            & 2.339   ${\pm}_{-0.343}^{+0.492}$ & 0.626 ${\pm}_{0.072}^{0.054}$ & 0.917 ${\pm}_{0.218}^{0.354}$                     \\
XMMXCS J095951.2+014045.8 & 149.963         & +1.679          & 0.372  & -24.057            & 2.128   ${\pm}_{-0.192}^{+0.238}$ & 0.557 ${\pm}_{0.035}^{0.029}$ & 0.734 ${\pm}_{0.110}^{0.146}$                     \\
XMMXCS J022401.9-050528.4 & 36.008          & -5.091          & 0.324  & -23.206            & 1.759   ${\pm}_{-0.364}^{+0.576}$ & 0.515 ${\pm}_{0.090}^{0.064}$ & 0.544 ${\pm}_{0.178}^{0.339}$                     \\
XMMXCS J095924.7+014614.1 & 149.853         & +1.770          & 0.124  & -22.717            & 1.252   ${\pm}_{-0.098}^{+0.113}$ & 0.472 ${\pm}_{0.024}^{0.022}$ & 0.339 ${\pm}_{0.044}^{0.054}$                    
\end{tabular}
\label{clusparams_xcs}
\end{table*}
\begin{figure*}
\centering
	\includegraphics[width=18cm,height=18cm,keepaspectratio,trim={2cm 5cm 5cm 2cm},clip]{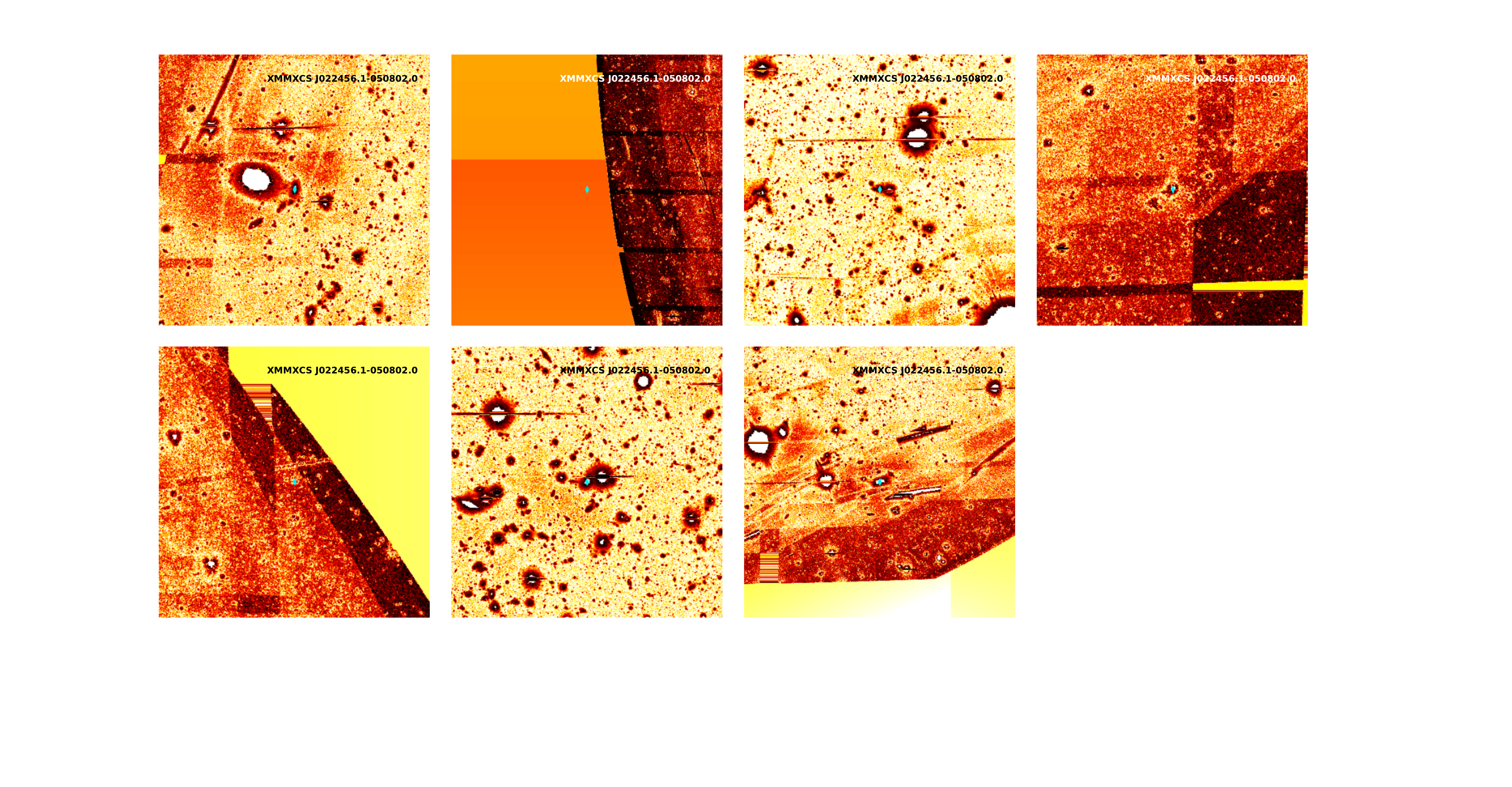}
    \caption{The 7 clusters ($1.5 \times 1.5$ Mpc on each frame) omitted from the sample due to poor photometry or bright source contamination. BCGs, if present on the frame, are marked with cyan diamonds. The images have been log-scaled and Gaussian-smoothed to show structure.}
    \label{fig:bcg_bad}
\end{figure*}
\par For each of the fields, $i$-band image data were then downloaded as cutouts (see Section \ref{hsc}) using a field size equivalent to $1.5 \times 1.5$ Mpc at the spectroscopic redshift of the cluster. These were checked against the value used here as a proxy for cluster radius (see Section \ref{quant_icl}) to ensure that the field size encompassed the size of the cluster as estimated by X-rays. The quality of the individual fields were checked at this stage, with 7 being discarded due to bright foreground source contamination, or being at the edge of a field. The 7 clusters with rejected image data are shown in Figure \ref{fig:bcg_bad}. Another 4 clusters were also rejected a posteriori, as they were agreed by the collaboration to be poor candidates. Our final sample therefore consists of 18 clusters (see Figure \ref{fig:all_clusters}); the corresponding $M_{\mathrm{X,500}}-z$ relation (see Section \ref{quant_icl}) can be seen in Figure \ref{fig:lx_z}. The clusters span a wide range in both redshift ($0.06<z< 0.5$) and halo mass ($10^{12.5} < M_{X,500} < 10^{14.5}$).
\par From the X-ray measurements, we estimate $R_{\mathrm{X, 500}}$ and  $M_{\mathrm{X, 500}}$ using the X-ray temperatures of the remaining clusters in our sample (the subscript $X,500$ referring to the value being derived from X-rays). The $R_{\mathrm{X, 500}}$ act as a proxy for the cluster radius and are used as physically-motivated aperture sizes for measuring ICL;  $R_{X,500}$ also has the benefit of lower levels of contamination from the background compared with larger cluster radii (e.g. $R_{200}$). We do, however, recognise that there is a significant caveat with this method, in that we are assuming the BCG to be a proxy for the centre of the cluster. While this is generally a reasonable assumption at low redshift (e.g. \citealt{2004ApJ...617..879L}), at higher redshift, there are an increasing number of clusters out of dynamical relaxation (e.g. \citealt{2011MNRAS.410.1537H}) with multiple BCG candidates; this may be resolved in future studies with deeper photometric coverage (e.g. mass-weighted centroid estimation via weak lensing).
\par Both $R_{\mathrm{X, 500}}$ and  $M_{\mathrm{X, 500}}$ are computed via the scaling relations of \cite{2005A&A...441..893A}, modelled as power laws:
\begin{equation}
    E(z)R_{X,500} = 1.104 \bigg[ \frac{\mathrm{k}T}{5 \ \mathrm{keV}}\bigg]^{0.57} \ \mathrm{Mpc} \ ,
\end{equation}
\begin{equation}
    E(z)M_{X,500} = 3.84  \times 10^{14} \bigg[ \frac{\mathrm{k}T}{5 \ \mathrm{keV}}\bigg]^{1.71} \ \mathrm{M_{\odot}} \ ,
\end{equation}
where $T_{X, 500}$ is the X-ray temperature (K) and $E(z)$ here is:
\begin{equation}
    E(z) = [{\Omega_{M}}(1 + z)^{3} + \Omega_{\Lambda}]^{-1/2} ,
\end{equation}
where $z$ is the cluster redshift and $\Omega_{\mathrm{M}}$ and $\Omega_{\Lambda}$ are our concordance cosmology values. The range of $R_{X,500}$ and $M_{X,500}$ values for our clusters are summarised in Table \ref{clusparams_xcs}. Although we recognise that the relation from \cite{2005A&A...441..893A} is derived from relaxed clusters (which may not be the case here), a recent paper from \cite{2017MNRAS.465..858G} investigated the luminosity-mass relation using statistically-complete Chandra data with masses derived via a hydrostatic mass analysis. They found no significant differences between relaxed and non-relaxed clusters when comparing masses derived from a $Y_{x}$-mass relation.
\section{Analysis}
\subsection{Background Over-subtraction - The `Divot Correction' Method}\label{section:divcorr}
A major concern regarding the measurement of ICL is not only the \textit{addition} of flux from excess sources (as discussed in the prior subsection and in Section \ref{image_depth}) but also, the \textit{over-subtraction} of flux. For space-based telescopes with low levels of background, this is generally less of a concern (e.g. HST); in the case of ground based telescopes, however, it provides a significant challenge for LSB science. For extended objects such as galaxies, issues arise due to modern commonly-used background estimation methods -  namely, spline-mesh approaches. Within the galaxy-modelling literature, this issue is long-known (e.g. \citealt{2015MNRAS.448.2530Z} and references therein); namely, that such approaches produce a `dearth' of flux around extended sources, termed here as a `divot'.
\par Divots occur because we are limited in our background estimation by the size of our chosen mesh, as we cannot accommodate for the wide range of angular extents of all objects in a frame. Hence, some light in extended object profile wings is often mistaken for background flux and mistakenly subtracted with the sky. Even in surveys such as HSC where background estimation is (more-or-less) state-of-the-art, these features still occur (see Figure \ref{fig:divot_demo}). This effect is doubly serious in the case of cluster and ICL science compared with isolated galaxies, as there is often a high source density (i.e. overlapping profile wings) which makes it nearly impossible to select a globally appropriate mesh size.
\par In an upcoming paper (Kelvin et al., in prep; Lee Kelvin, priv. comm.), we attempt to address these problems, providing survey comparisons and suggesting potential solutions. To do so, we have produced a pipeline to correct for such flux over-subtraction effects. We acknowledge that post-processing is less preferable than an optimised survey strategy, especially given that our method involves parametric estimates that we attempt to avoid as much as possible when measuring our ICL values (Section \ref{quant_icl}). In this case (and in many others) however, this is not an option for either past or present surveys that have not prioritised LSB science in their observational approach. The construction, application and limits of the aforementioned pipeline will be the subject of a separate paper; here, we instead provide an abridged description of its operation and use in the context of this work.
\par The pipeline, which is written in \textsc{R} and is primarily \textsc{SExtractor} (version 2.19.5) and \textsc{SWARP} based, operates on an image in three major steps as follows:
\begin{enumerate} 
\item \textbf{Object detection/modelling:} firstly, \textsc{SExtractor} is run on a given input image. The settings used are similar to those used in \cite{2018MNRAS.478.4952F}. Since \textsc{SExtractor} version 2.8 (e.g. \citealt{2009MmSAI..80..422B}), it is possible to fit models to the light profiles of objects detected by the algorithm. There are several model types available (e.g. delta function, Ferrer profile, exponential profile, S\'{e}rsic profile). Here, we opt for a single-S\'{e}rsic model (see equation \ref{eq:1}). All detected objects in the frame are modelled with a S\'{e}rsic profile, which are fit through a Levenberg-Marquardt ${\chi}^{2}$ minimization algorithm. The S\'{e}rsic profile has the following form:
\begin{equation} \label{eq:1}
I(R) \ = \ I_{e}\exp{ \{ b_{n}[(R/R_{e})^{1/n} \ - \ 1] \}} \ ,
\end{equation}
where $I(R)$ is the intensity of an object at radius $R$, $R_{e}$ is the effective radius, $I_{e}$ is the object intensity at the effective radius, $n$ is the S\'{e}rsic index and $b_{n}$ is a product of incomplete gamma functions as described in \cite{1991A&A...249...99C}. \textsc{SExtractor} imposes an internal hard limit on the range of S\'{e}rsic indices ($0.5 < n < 8$); the majority in this work fall around $0.5 < n < 4$. The result of doing so is an image frame containing the modelled light profiles of all catalogue objects.
\item \textbf{Differential inversion:} in order to estimate the flux loss in object profile wings caused during the image processing stage, we then take the difference between the input image and the image containing the object models. The result is then inverted, creating the `divot correction' (see centre panel of Figure \ref{fig:divot_demo}).
\begin{figure*}
\centering
	\includegraphics[width=18cm,height=18cm,keepaspectratio, trim={3.5cm 6cm 2cm 5cm},clip]{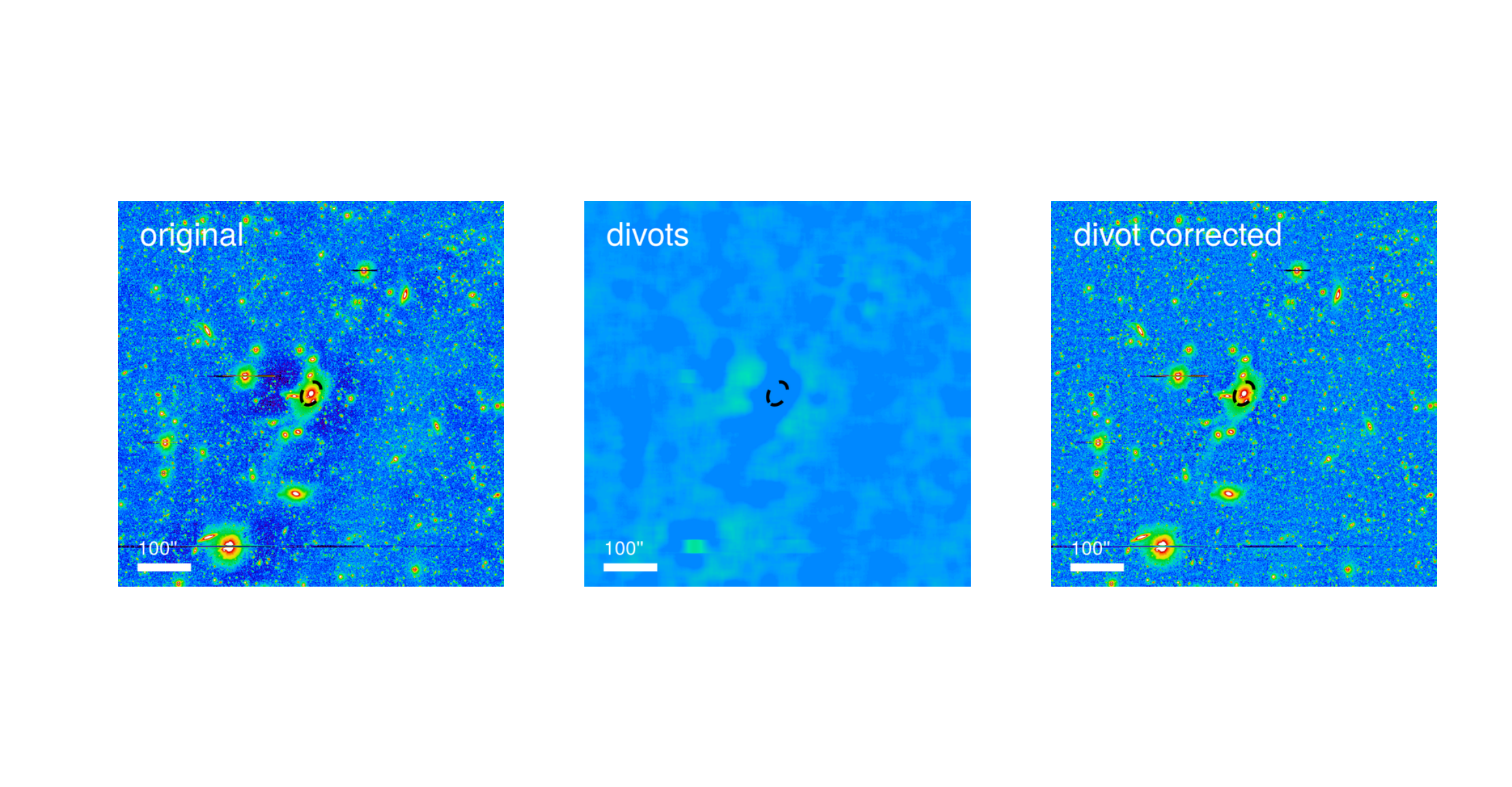}
    \caption{An example of the `divot correction' method used in this work (shown here for cluster XMMXCS J232923.6-004854.7, $z = 0.3$); all images have the exact same scaling (from Kelvin et al., in prep). The first image depicts the data prior to correction; as is visible, there is a dearth of flux in the regions around the BCG and its satellites. The second shows the estimated divot correction (the divot corrections are smoothed using a 5-pixel FWHM Gaussian kernel); the third shows the resultant image after implementation. As is visible, there is a vast improvement, with the sky level varying far more smoothly.}
    \label{fig:divot_demo}
\end{figure*}
\item \textbf{Coaddition:} the divot correction image is then added on to the original image using \textsc{SWARP} (\textsc{LANCZOS3} interpolation; this was selected as recommended in the \textsc{SWARP} user manual, but as the resolution of the images is identical, no resampling is necessary), thus providing an approximate flux `correction' (see Figure \ref{fig:divot_demo}).
\end{enumerate}
\par There is an obvious caveat in our approach, namely with our selection of a single-S\'{e}rsic profile with which we fit to all galaxies in a frame. We therefore assume that object wings will follow those of a S\'{e}rsic profile. This estimate is often cuspier than, for example, the true profiles of BCGs, of which some are thought to be multi-component objects (see, e.g. \citealt{2014MNRAS.443..874B}, \citealt{2015MNRAS.448.2530Z}, \citealt{2019ApJ...874..165Z}; \citealt{2016ApJ...820...42I} see also Section \ref{resultsiv}) and may, for example, lead to residuals which are added into the image which are not part of the divot. As well as this, there will be some dependence on the reliability of the correction with both cluster extent and redshift; with that said, we choose large postage stamps (in excess of $R_{X,500}$ in most cases) when modelling the divots (to provide a sense of scale, the objects modelled here range from $\sim 2^{\prime\prime}-13^{\prime\prime}$). Although we appreciate the simplicity of this approach, the addition of other components to hundreds of models (as well as attempting to accurately morphologically classify all detected objects in a frame), provides not only significant computational cost challenges, but also adds additional free parameters which may not be necessary for all objects and may lead to less reliable fits (see arguments in \citealt{2018MNRAS.478.4952F}). We therefore instead caution the reader that our estimates represent, most likely, a lower-limit estimate on the true value of the total wing flux loss during processing. The divot method allows us to quantify the over-subtraction but is not a substitute for a full pipeline sky-subtraction reduction analysis.
\subsection{BCG Photometry}\label{bcg:photo}
\par We apply three methods of quantifying the flux contribution from our cluster BCGs: total flux within an aperture of radius 50 kpc (e.g. \citealt{2008MNRAS.387.1253W}) or two parametric models: a single, free S\'{e}rsic fit, or a de Vaucouleurs model with a fixed S\'{e}rsic index of 4. We choose an aperture of radius 50 kpc primarily as other authors have found that this radius corresponds approximately to the region where there is an excess of light in BCGs compared with a de Vaucouleurs profile (e.g. \citealt{2014A&A...565A.126P}). We prefer, given the nature of our data, to take a simplistic approach over attempting to fit multiple components here. We take a similar approach as in our previous work in this respect (\citealt{2018MNRAS.478.4952F}), where we assessed the performance of the pipeline for SDSS data. There are numerous arguments as to the best model to fit; most notably, a two-component model which includes the addition of an exponential halo to a S\'{e}rsic profile (e.g. \citealt{2011ApJS..195...15D}; \citealt{2015MNRAS.448.2530Z}; \citealt{2013MNRAS.436..697B}). However, we take the approach in this work that disentangling the BCG from the ICL is non-trivial to achieve, given how much they are closely linked in terms of evolutionary history (e.g. \citealt{2012MNRAS.425.2058B}; \citealt{2018ApJ...864..149S}; \citealt{2016ApJ...820...42I}), so include parametric model fits primarily as a comparative measure. For our results, due to them being non-parametric, we use the aperture values to represent our BCG fluxes.
\par
\begin{figure*}
\centering
	\includegraphics[width=18cm,height=18cm,keepaspectratio,trim={0cm 0cm 1cm 0cm},clip]{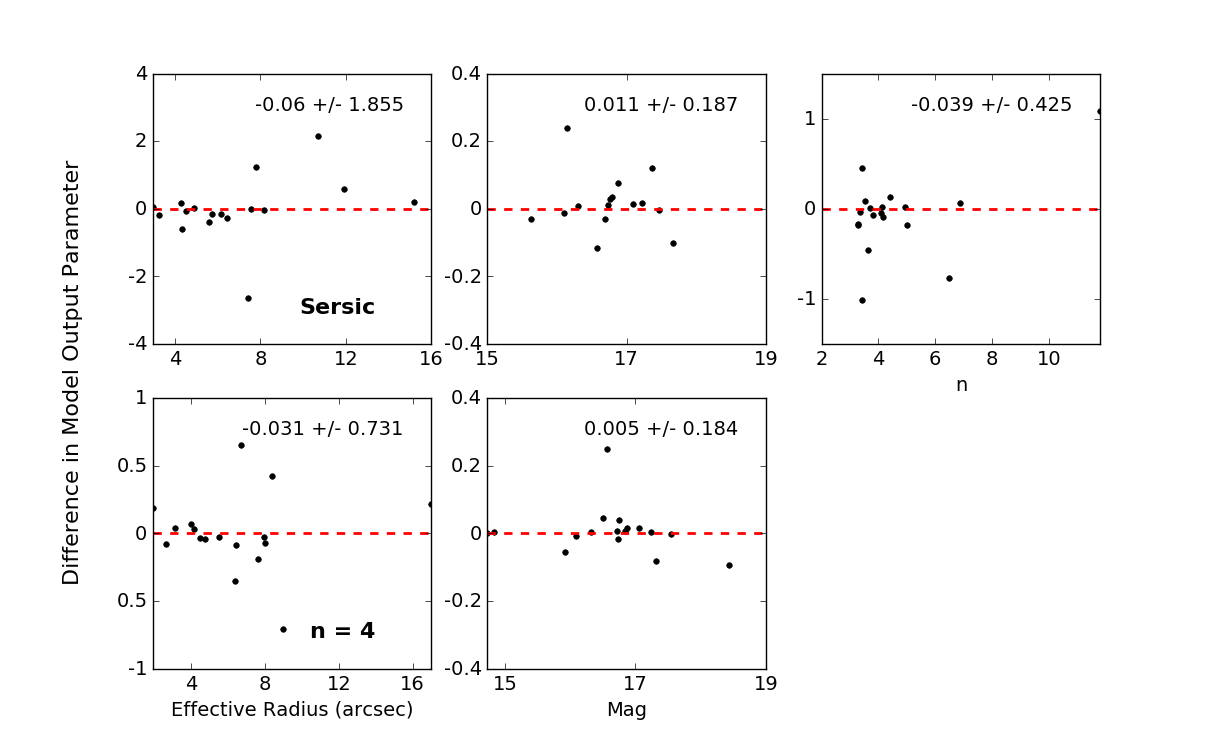}
    \caption{Differential comparison between the input and output parameters for the cluster BCGs in this sample, with and without an added divot correction. The values at the top of each frame represent the median deviation and $rms$. The top and bottom panels represent the outputs for a free S\'{e}rsic profile and a de Vaucouleurs profile respectively. The fit parameters for both the non-corrected and corrected cases tend to be reasonably similar and there are no clear biases present upon using a divot correction.}
    \label{fig:div_nodiv_sigma_outputs}
\end{figure*}

 We model our galaxies using the SIGMA pipeline (Structural Investigation of Galaxies via Model Analysis, see \citealt{2012MNRAS.421.1007K}), using a similar implementation as in \cite{2018MNRAS.478.4952F}. SIGMA is a software wrapper written in R that performs a full model fit of a given object using GALFIT 3 (see \citealt{2010AJ....139.2097P}), including an estimate of the field PSF using PSFEx (see \citealt{2013ascl.soft01001B}). The weight maps used in this procedure are those generated by the HSC pipeline. We fit the BCGs simultaneously with their brightest 3 neighbours, masking out their centres (to mitigate saturation issues) and the remaining objects in the field. We produce models for our objects pre- and post-divot correction (see Section \ref{section:divcorr}) and use the post-divot corrected models because of the correction to the profile wings of our objects. Generally, the output parameters are similar in both cases (see Figure \ref{fig:div_nodiv_sigma_outputs}), and do not show any obvious biases. 
\par It is important to mention that we do not use the PSFs generated by SIGMA when masking of stars on our images (e.g. to estimate the contamination extent); rather, their use is to provide a sufficiently well-approximated model for our BCG model fits. This is because the PSFs generated by SIGMA are not estimated out to large enough radii to account for the wings of the brightest stars on our frames ($\sim 0.2^{\prime}$). PSFEx is not optimised for the purposes of producing extended PSFs; indeed, using PSFs with a small angular extent both for the purpose of masking and removal of wings from point source contamination represent two of the most commonly-cited issues regarding the robustness of LSB photometric studies (e.g. \citealt{2015MNRAS.446..120D}, in the context of deep ATLAS-3D survey data; see also \citealt{1991ApJ...369...46U}, \citealt{2009PASP..121.1267S}, \citealt{2016ApJ...823..123T}, \citealt{2020A&A...644A..42R}, \citealt{2020MNRAS.491.5317I}). For a more detailed description of the masking process, see Section \ref{section:masking}.
\par
\begin{figure*}
\centering
	\includegraphics[width=18cm,height=18cm,keepaspectratio,trim={18cm 4cm 16cm 0cm},clip]{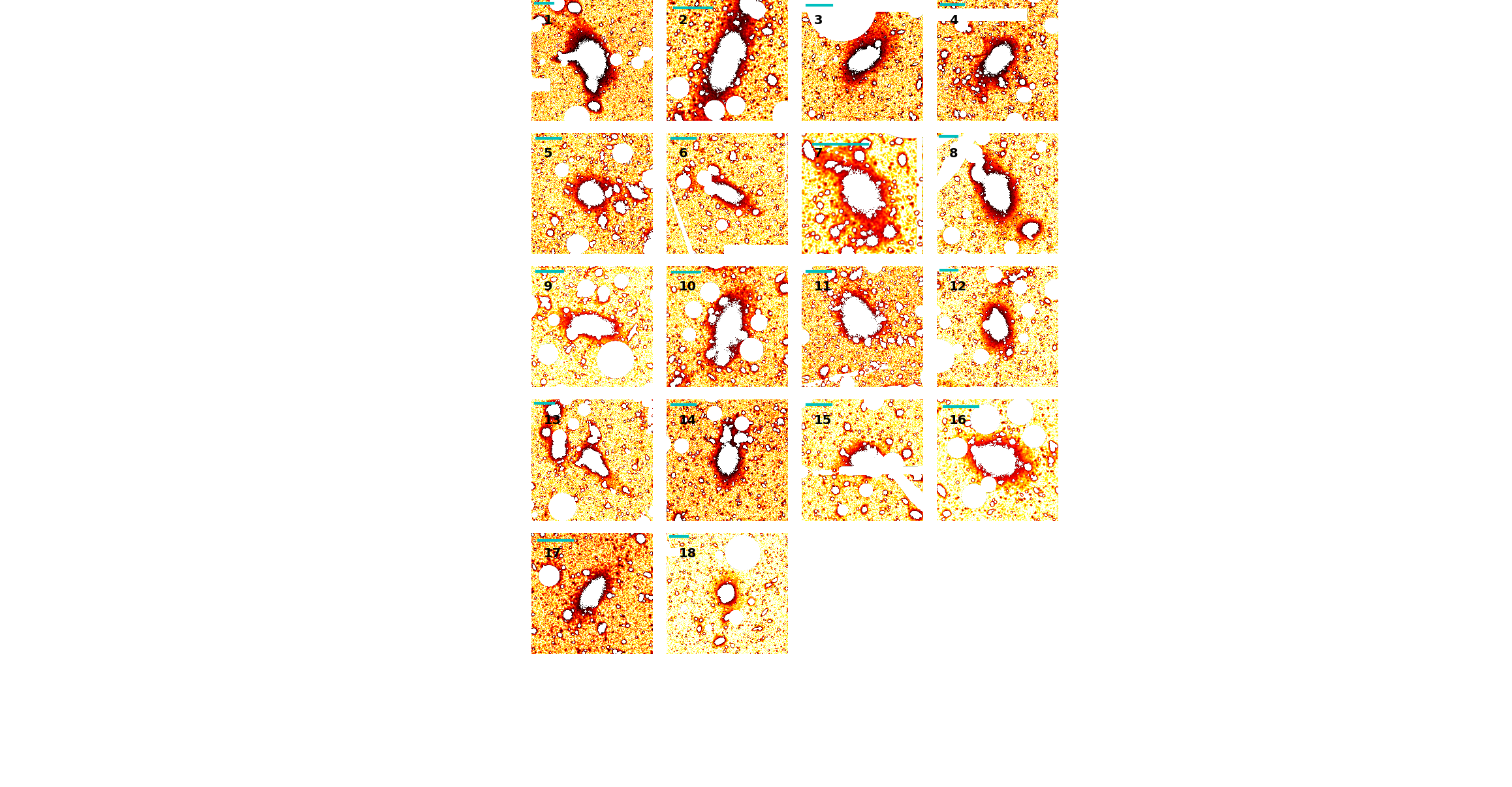}
    \caption{The final equivalent 25 $\mathrm{mag/arcsec^{2}}$ masked and divot-corrected images (numerical labels have been included for clarity in discussion), zoomed to 40\% of the $R_{X,500}$ value of each respective cluster centred on the BCG (with the exception of panel 7). The cyan line at the top left of each panel is the equivalent $30^{\prime\prime}$ pixel scale for each image. To show structure, the images have been log-scaled and smoothed.}
    \label{fig:all_clusters}
\end{figure*}

 In most cases, the three methods of quantifying BCG magnitude agree within a few percent, with the aperture values generally yielding slightly lower values due to there being no wing extrapolation (e.g. \citealt{2018MNRAS.478.4952F}). There are, however, a couple of cases where there is a disagreement between values of $\sim 10$\% or higher:
\begin{enumerate}
    \item \textbf{XMMXCS J095901.2+024740.4} the highest redshift system in this work ($z = 0.51$; panel 7 of Figure \ref{fig:all_clusters}), with the faintest BCG apparent magnitude from an integrated model ($m_{i} = 18.51$). The BCG flux fraction for this system with respect to the cluster within $R_{X,500}$ doubles using the best S\'{e}rsic fit over either the aperture or de Vaucouleurs values (0.34, compared with 0.17 and 0.21 respectively). From our work in \cite{2018MNRAS.478.4952F}, we found that galaxy models tended to degrade with decreasing surface brightness; indeed, of all of the BCGs modelled here, the S\'{e}rsic fit for this system has the largest relative error. 
    \item \textbf{XMMXCS J095951.2+014045.8:} closer inspection of the system using the DS9 software revealed it to be a cD-type (panel 14 of Figure \ref{fig:all_clusters}); this extra flux may potentially have been missed through using an aperture to measure the BCG (e.g. for a recent paper on the effect of cD haloes when fitting galaxies, see \citealt{2015MNRAS.453.4444Z}) and more heavily contributed to the ICL fraction, as both fitted models give a larger fraction of cluster light attributable to the BCG (0.28 in either case, compared with 0.19 for the aperture estimate). As aforementioned, such cases are testament to the caveats of a non-parametric approach.
\end{enumerate}
\subsection{Masking}\label{section:masking}
As in every photometric survey, HSC imaging is not free from artifacts. Although the processing algorithm has been optimally designed to avoid such defects wherever possible, some sources of excess flux remain. These include artifacts from overexposed stars, telescope ghosts, satellite trails and cirrus, to name a few (refer to \citealt{2015MNRAS.446..120D} for a comprehensive summary). This is shown in Figure \ref{fig:bcg_bad}, which constitutes examples of clusters in XCS which were not included in the final sample due to heavily contaminated photometry in HSC.
\par For our sample, we create custom masks in order to minimise the contribution to ICL flux from artifacts. Although the HSC pipeline does produce masks as output, we opt to generate our own as an attempt to more comprehensively remove artifacts, such as extended diffraction spikes from bright stars which are often not cleanly removed. We refer the reader to \cite{2018PASJ...70S...5B} for more details of the masking method used in the HSC pipeline.
\par For our custom masks, we begin with the binary masks generated by the HSC pipeline. The binary masks contain numerical identifiers in order to differentiate between different `layers' of the masks - namely artifacts/saturated stars vs. objects. From these, we generate our mask layers via 3 stages: 
\begin{enumerate}
\item \textbf{Bad Pixel Masking:} we begin by first identifying the `bad pixel' regions thresholded out by the HSC pipeline. These regions are then masked out, and constitute the first mask layer. These include: regions which have been incorrectly weighted by the weight maps, saturated pixels and some of the artifacts generated by bright stars. 
\item \textbf{Star Masking:} next, we run SExtractor across all of the images. We set a detection threshold for our objects at $10 \sigma$; with other parameters (such as saturation level, etc.) set to roughly the same values as those used during our running of SIGMA. We allow SExtractor to approximate a rough background level using a large mesh size to account for any extended bright sources (128 pixels). The purpose of this step is primarily to identify brighter, more compact objects within the frame, for which we do not require absolutely accurate photometry. 
\par For fainter stars, we query the Gaia DR2 catalogue (\citealt{2018A&A...616A...1G}) for both photometry and astrometry. The Gaia mission aims to collect both photometry and astrometry for $\sim 10^{9}$ stars in the Milky Way (for science objectives, see \citealt{2016A&A...595A...1G}). We produce catalogues of stars within the frames of our images, and mask stars out with $17 < G < 21$ (mean apparent magnitude value in the $G$-band from Gaia, see technical paper for the filter curve: \citealt{2010A&A...523A..48J}; $G \sim 21$ is the survey limit). There is around a 10\% rate of contamination in Gaia by elliptical galaxies; we follow the prescription outlined in \cite{2017MNRAS.470.2702K} and apply a cut using the `astrometric excess noise' parameter, $ans$ ($\mathrm{log_{10}}(ans) < 0.15(G - 15) + 0.25$), which they found to be $\sim$ 95\% effective; upon visual inspection, none of the BCGs were masked in this way. We then apply the following empirical masking formula used canonically in HSC\footnote{\url{https://hsc-release.mtk.nao.ac.jp/doc/index.php/bright-star-masks/}} to define our exclusion apertures:
\begin{equation}r = {A_{0}} \times 10^{B_{0} (C_{0} - i)} + A_{1} \times 10^{B_{1} (C_{1} - i)}
\end{equation}
where $r$ is in pixels, $i$ are the HSC $i$-band magnitudes as measured by SExtractor (Kron aperture, \citealt{1980ApJS...43..305K}) and  $A_{0} = 200$, $B_{0} = 0.25$, $C_{0} = 7.0$, $A_{1} = 12.0$, $B_{1} = 0.05$, and $C_{1} = 16.0$.
\par For brighter stars ($G$ $<$ 17 in our case), this approach is not recommended. Although some bright stars are masked in HSC already, there are many missing due to the prior use of the much less complete NOMAD survey (\citealt{2004AAS...205.4815Z}) compared with the Gaia survey, which will be used for future releases as detailed in \cite{2018PASJ...70S...7C}. Instead, we create custom masks across all frames by hand for the brightest stars, any other point-like sources missed in our catalogues from Gaia and any visible diffraction spikes (a similar method to that used, for example, in \citealt{2018MNRAS.474..917M} and \citealt{2015MNRAS.449.2353B}). Using the same method, we also mask out all non-cluster galaxies brighter than the BCG via careful visual inspection of the cluster field, following \cite{2015MNRAS.449.2353B}. We used the SAO DS9 imaging software to view our images, which includes an array of tools for image visualisation ideal for these purposes, including optimised Gaussian smoothing kernels and high contrast scaling (useful for scaling masks to accommodate stellar wings). Masks were then created by hand using the region definition tool in DS9, and subsequently converted to fits format using the open-source \textsc{mkmask} software (courtesy: Rolf Janssen). 
\begin{table}
\centering
\caption{The $k$-correction ($k_{i,B}$), cosmological dimming and equivalent $B$-band surface brightness limits at which we observe (where $\mu_{i,obs}$ is equivalent to $\mu_{B,rest} = 25$) our clusters, used to generate isophotal masks.}
\begin{tabular}{l|l|l|l}
XCS ID                    & $k_{i,B}(z)$ & $2.5 \mathrm{log_{10}}(1+z)^{4}$ & ${\mu}_{i,obs}$ \\ \hline
XMMXCS J022456.1-050802.0 & -1.566                 & 0.350       & 23.784                \\
XMMXCS J161039.2+540604.0 & -1.304                 & 1.263       & 24.959                \\
XMMXCS J233137.8+000735.0 & -1.428                 & 0.877       & 24.450                \\
XMMXCS J232923.6-004854.7 & -1.350                 & 1.142       & 24.792                \\
XMMXCS J161134.1+541640.5 & -1.304                 & 1.263       & 24.958                \\
XMMXCS J095902.7+025544.9 & -1.289                 & 1.299       & 25.010                \\
XMMXCS J095901.2+024740.4 & -1.105                 & 1.741       & 25.635                \\
XMMXCS J100141.6+022538.8 & -1.523                 & 0.508       & 23.985                \\
XMMXCS J095737.1+023428.9 & -1.257                 & 1.378       & 25.121                \\
XMMXCS J022156.8-054521.9 & -1.392                 & 1.001       & 24.608                \\
XMMXCS J022148.1-034608.0 & -1.183                 & 1.556       & 25.374                \\
XMMXCS J022530.8-041421.1 & -1.498                 & 0.580       & 24.082                \\
XMMXCS J100047.3+013927.8 & -1.430                 & 0.865       & 24.435                \\
XMMXCS J022726.5-043207.1 & -1.341                 & 1.168       & 24.827                \\
XMMXCS J022524.8-044043.4 & -1.387                 & 1.018       & 24.631                \\
XMMXCS J095951.2+014045.8 & -1.259                 & 1.374       & 25.115                \\
XMMXCS J022401.9-050528.4 & -1.324                 & 1.218       & 24.894                \\
XMMXCS J095924.7+014614.1 & -1.525                 & 0.500       & 23.975               
\end{tabular}
\label{t:kcorrs_ii}
\end{table}
\item \textbf{Isophotal Mask Creation}: we then produce isophotal masks for each of our frames (see discussion in \ref{quant_icl}), below which we define the ICL to be measured and apply these in conjunction with our bad pixel and star masks when performing photometry. To do so, we use an effective surface brightness detection threshold in the rest-frame of 25 mag/$\mathrm{arcsec^{2}}$ (an approach similar to that carried out on the CLASH cluster sample by \citealt{2015MNRAS.449.2353B}). To compare our results with \cite{2015MNRAS.449.2353B}, we also shift our equivalent surface brightness threshold at which we measure ICL to that of the rest-frame $B$-band.
For the $B$-band equivalent threshold, we introduce the following equation:
\begin{equation}
    {\mu}_{i,obs} = {\mu}_{B,rest} + 2.5 \mathrm{log_{10}}(1 + z)^{4} + k_{i,B}(z) \ ,
\end{equation}
where ${\mu}_{i,obs}$ is the limit at which we observe, ${\mu}_{B,rest}$ is the equivalent rest-frame surface brightness in the $B$-band, $2.5\mathrm{log_{10}}(1 + z)^{4}$ is the bolometric cosmological surface brightness dimming term and $k_{i,B}(z)$ is the $k$-correction term, defined here as:
\begin{equation}
    {k}_{i,B}(z) = M_{i,obs}(z) - M_{B,rest}(z) \ ,
\end{equation}
where $M_{i,obs}(z)$ and $M_{B,rest}(z)$ are the pseudo-absolute magnitudes derived for each respective waveband at a given redshift for our choice of stellar population synthesis model. These are computed via the EZGAL software (\citealt{2012PASP..124..606M}), assuming an old stellar population with a formation redshift of $z_{f} = 3$, solar metallicity ($Z_{\odot}$) and passive evolution thereafter, using the models of \cite{2003MNRAS.344.1000B} coupled with a \cite{2003PASP..115..763C} IMF (also resembling the methodology of \citealt{2018MNRAS.474.3009D}). We list our $B$-band limits in Table \ref{t:kcorrs_ii}.
\par While we appreciate that it is unlikely that the stars contained within our BCGs evolved entirely in situ, most BCGs have shown little evidence for significant growth through starburst activity at $z < 1$ and are primarily assumed to gain mass through mergers with satellites containing reasonably similar stellar populations (or even more passive, e.g. \citealt{2009MNRAS.398.1129G}), so we consider this assumed `burst' model reasonable for simplicity (this was an assumption also made by \citealt{2015MNRAS.449.2353B}). There are, however, an increasing number of studies showing a younger age for the ICL component compared to the BCG (see, for example, \citealt{2014ApJ...794..137M}; \citealt{2018MNRAS.474..917M}; \citealt{2017ApJ...846..139M}; \citealt{2019A&A...622A.183J}). As an aside, we also performed an additional check to ensure that correcting to the $B$-band did not result in any serious biases from the rapid fade of bluer stellar populations with redshift (see Appendix). 
\par We show how the choice of metallacity and formation redshift affects our $k_{i,B}(z)$ values in the Appendix, for the mean values of our sample split in two bins about the mean redshift ($0 < z < 0.28$ and $0.28 < z < 0.5$ respectively); in short, there is an $rms$ of $\pm 0.3$ magnitudes in $k_{i,B}(z)$, depending on the model of choice. Interpolating from Figure \ref{fig:isofrac} (see upcoming discussion), this translates at 25 $\mathrm{mag/arcsec^{2}}$ to a difference in $\pm$5\% of the final ICL value.
\end{enumerate}
\subsection{Quantifying the Systematic Background}\label{image_depth}
In all astronomical image data, a systematic background exists. At visible wavelengths, it is partially caused by faint galaxies below the survey limit (which is a caveat to our method, refer to discussion below), the wings of bright sources such as stars or contaminant galaxies and residual flux from the sky (e.g. \citealt{10.1111/j.1365-2966.2009.14739.x}). In order to better understand this in the context of our image data, we performed a test by applying photometry on injected mock profiles so that we could trace the additional flux contribution at a given surface brightness. We performed this test on `control' frames offset from each of the clusters in this study. The 18 control frames selected were patches of sky within the HSC-SSP footprint, offset at random by $0.5^{\circ}$ from the centre of the original frames. We chose to use representative control frames so as to prevent any contributions from ICL that may be present. The control frames were subject to an identical masking method as that used in the cluster frames, were weighted using the HSC-generated weight maps (inverse variance) and were not divot-corrected. 
\par For each of the frames, 10 random positions were selected. To mimic an ICL-like profile (found by numerous authors to be approximately exponential, e.g. \citealt{2005ApJ...618..195G}; \citealt{2007MNRAS.378.1575S}; \citealt{2019ApJ...874..165Z}), we generated an exponential model ($n = 1$, $R_{e} = \ <{R_{X,500}>}/4$, $\theta = 50^{\mathrm{\circ}}$, $a/b = 0.8$; S\'{e}rsic index, effective radius, position angle and axis ratio respectively) at 9 surface brightness levels (24 - 28 $\mathrm{mag/arcsec^{2}}$ in steps of $0.5\  \mathrm{mag/arcsec^{2}}$; where, for reference, the faintest limit is $\sim$3 magnitudes below that which we measure for the ICL in our clusters). The profiles were convolved with the field PSF from SIGMA (as per the modelling process for the BCGs) and an idealised Poisson noise component was added. We injected the models at 10 random positions within each of the control frames, measuring the difference between the input and output flux values using a fixed circular aperture equivalent to the selected effective radius of our models ($\sim 2000$ models in total, for a similar method, see \citealt{2012MNRAS.425.2058B}). 
\par
\begin{figure*}
\centering
	\includegraphics[width=19cm,height=19cm,keepaspectratio,trim={0cm 2cm 0cm 1cm},clip]{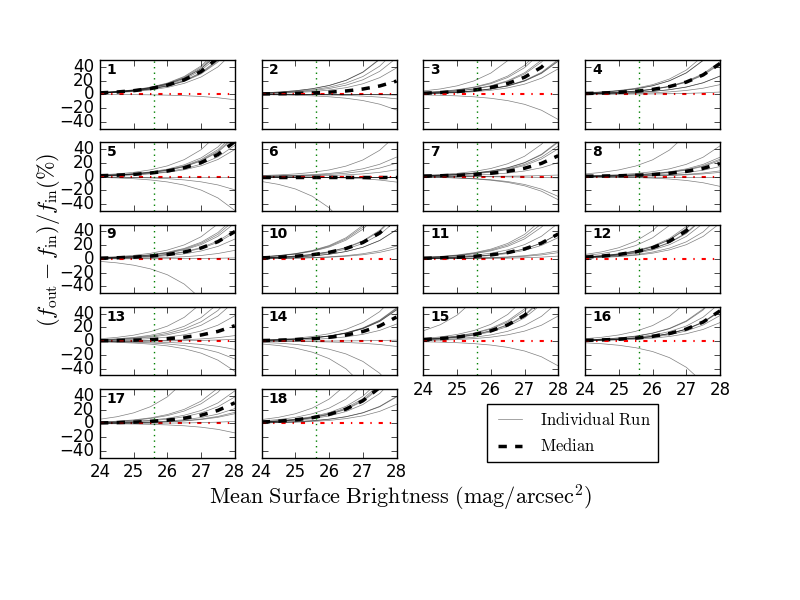}
    \caption{The results from performing photometry on $\sim 2000$ mock ICL profiles injected into HSC data (without divot corrections; numerical labels have been included as in Figure \ref{fig:all_clusters}). Each frame in the subplot represents a given control field. The plot shows the relative percentage deviation in flux ($(f_{\mathrm{out}} - f_{\mathrm{in}})/f_{\mathrm{in}}$), where $f_{\mathrm{in}}$ is the raw mock profile flux measurement and $f_{\mathrm{out}}$ is the flux measurement of the profile after implantation in a HSC control frame for a given `ICL-like' profile (see text) with respect to mean surface brightness (the average surface brightness across a mock profile). The green dotted lines show the isophote of lowest surface brightness used in this work.}
    \label{fig:sep_runs}
\end{figure*}

\begin{figure}
\centering
	\includegraphics[width=\columnwidth,keepaspectratio,trim={0cm 0cm 0cm 0cm},clip]{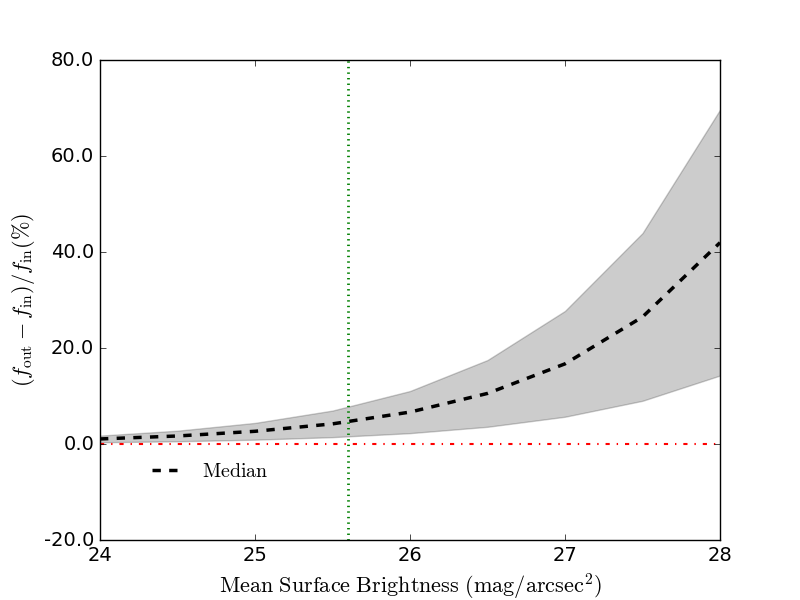}
    \caption{Stack of median deviations of recovered profiles across all frames with respect to mean surface brightness. The grey shaded region indicates the 1$\sigma$ scatter. The green line shows the isophote of lowest surface brightness used in this work.}
    \label{fig:stacky2}
\end{figure}

 Figure \ref{fig:sep_runs} shows the bulk output across the fields, with Figure \ref{fig:stacky2} showing the stacked median for all of the control frames. From our mock photometry, we detect a $<$5\% excess of the input flux on average for an ICL-like profile over the range of our $B$-band equivalent surface brightness levels (23.74 - 25.64 $\mathrm{mag/arcsec^{2}}$). There is obvious scatter on a case by-case basis (for example, panel 16 of Figure \ref{fig:sep_runs}, see also panel 16 of Figure \ref{fig:all_clusters}); from eyeballing, the predominant cause of this seems to be due to source-heavy frames (e.g. many/clustered sources or bright sources such as stars present). Moreover, we will show evidence in Section \ref{resultsiv} that the flux lost through the divot effect at the range of isophotal levels at which we measure the ICL is approximately $4 \times$ the background contribution; hence, we do not correct for it here (see further discussion of systematics in Section \ref{quant_icl}). It is again, however, worth noting that this method does not quantify the flux contribution of the population of faint galaxies below the survey limit (indeed, it is an issue with all similar observational studies of ICL, e.g. see \citealt{2019ApJ...874..165Z}).
\subsection{Quantifying ICL}
\label{quant_icl} 
\par Observationally, past studies have generally taken two approaches when quantifying the amount of ICL present in a cluster: a parametric approach using model fitting (e.g. \citealt{2005ApJ...618..195G, 2007ApJ...666..147G, 2013ApJ...778...14G}; \citealt{2017ApJ...846..139M}), or by summing up the contribution of ICL below a set (usually isophotal) limit while masking out the BCG and any satellites (e.g. \citealt{2012MNRAS.425.2058B, 2015MNRAS.449.2353B}; \citealt{2005AAS...20717714K, 2007AJ....134..466K}; \citealt{2018MNRAS.474..917M}). Other approaches looking at either the shape of the BCG+ICL profile (see upcoming discussion) or the so-called `colour profile' (namely, how the colour of the ICL spatially varies across the cluster) have also measured the flux in isophotes or annuli to acquire a 1-D profile (e.g. \citealt{2019ApJ...874..165Z}; \citealt{2018MNRAS.474.3009D}; \citealt{2012MNRAS.425.2058B}). 
\par When modelling a profile, one must assume a prior; exactly the best model to use when describing the BCG+ICL profile varies enormously across studies, with some recommending a double de Vaucouleurs profile (e.g. \citealt{2007AJ....134..466K}), some using a S\'{e}rsic+Exponential (e.g. \citealt{2007ApJ...664..226L}) and others more complicated models still (e.g. \citealt{2019ApJ...874..165Z}). Choosing the wrong profile can lead to large uncertainties (e.g. \citealt{2015MNRAS.448.2530Z}); as well as this, the degeneracies present when using multiple component fits mean that one cannot readily disentangle individual flux contributions without dynamical information (e.g. \citealt{2010MNRAS.405.1544D}). As per our masking methodology outlined in Section \ref{section:masking}, we take an isophotal approach to measuring the ICL in our clusters, which we do for two reasons: simplicity, and to keep our assumptions minimal. While the approach of using a surface brightness limit is not perfect (and often leads, according to \citealt{2011ApJ...732...48R}, to a lower ICL estimate), it is at least model independent. Here, we choose a limit of ${\mu}_{B} = 25 \ \mathrm{mag/arcsec^{2}}$ in the rest frame $B$-band, similar to \cite{2015MNRAS.449.2353B}; we discuss our methodology in Section \ref{section:masking}.
\par After applying a mask (which includes an isophotal threshold), we sum the weighted flux within an aperture of $R_{X, 500}$ centred on the cluster BCG and repeat the process without an isophotal limit (Section \ref{section:masking}). We also provide comparisons at the equivalent surface brightness levels of 24 and 26 $\mathrm{mag/arcsec^{2}}$ respectively to assess the effect of changing the selected surface brightness on the recovered ICL. The ICL measurement errors, $E(\mathrm{ICL})$, are computed directly from the image variances as follows: 
\begin{equation}
E(\mathrm{ICL}) = \sqrt{\bigg(\frac{\sigma_{ICL}}{f_{tot}} \bigg)^{2} + \bigg(\frac{f_{ICL} \times \sigma_{tot}}{f_{tot}^2} \bigg)^{2}} \ ,
\end{equation}
where the subscripts `$ICL$' and `$tot$' refer to the ICL and total flux respectively, $f$ is the flux in counts and $\sigma$ denotes the standard deviation.
\section{Results}\label{resultsiv}
\subsection{How Much of a Cluster is ICL?}
\par For comparison, we measure the ICL for our clusters before and after applying a divot correction. The measurements are summarised in Table \ref{tab:icl_amounts}. In Section \ref{sec:iclbcg}, we will provide more extensive comments on our results and their consequences for BCG evolution; here, we restrict our commentary towards the inferred \textit{systematics} involved in ICL measurement for ease of comprehension.
\begin{figure}
	\includegraphics[width=\columnwidth]{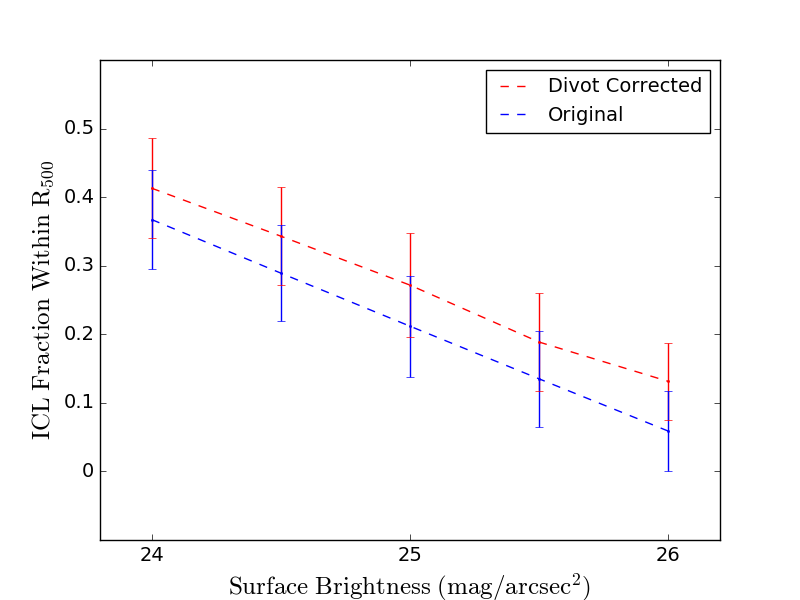}
    \caption{The stacked ICL fraction at a selection of equivalent surface brightnesses, comparing the divot corrected and uncorrected cases. The errorbars depict the 1-$\sigma$ scatter across all clusters in the sample.}
    \label{fig:isofrac}
\end{figure}
\par For our clusters, with the inclusion of a divot correction, the mean ICL contribution to the total cluster light at ${\mu}_{B,rest} = 25 \ \mathrm{mag/arcsec^{2}}$ sits at around $24 \%$. It is immediately clear from Table \ref{tab:icl_amounts} that applying a divot correction has a significant effect on the overall recovered value for the ICL ($\Delta f$ being the difference in ICL to total cluster light between the divot corrected and uncorrected values); Figure \ref{fig:isofrac} illustrates this difference, for equivalent surface brightness limits in $B$ from $24-26 \ \mathrm{mag/arcsec^{2}}$ in steps of 0.5. On average, the ICL fraction is $\sim 5 \%$ higher with a divot correction included, which represents $\sim$ 20\% of the mean measured ICL light fraction overall. The final masked, divot-corrected images are shown in Figure \ref{fig:all_clusters}.
\par Our results illustrate exactly how crucial it is to account for the flux over-subtraction problem around objects in surveys. As stated previously, because the divot corrections are modelled with a `one-size-fits-all' S\'{e}rsic profile, it is likely that the `true' net flux loss is underestimated due to our choice of S\'{e}rsic profile with which to model our divot corrections, with $\sim$ 50\% of BCGs tending to have an additional `halo' as well as a central bulge by $z < 0.1$ which is debated to be either ICL or a BCG component (\citealt{2015MNRAS.448.2530Z}). In addition, our method, as for other observational methods for measuring ICL which utilise a surface brightness limit, clearly cannot account for ICL in projection of the BCG. Whilst we appreciate that there is a method dependency when measuring ICL, there is still a significant difference upon inclusion of a divot correction when changing the surface brightness limit (Figure \ref{fig:isofrac}). 
\begin{figure}
    \centering
	\includegraphics[width=\columnwidth]{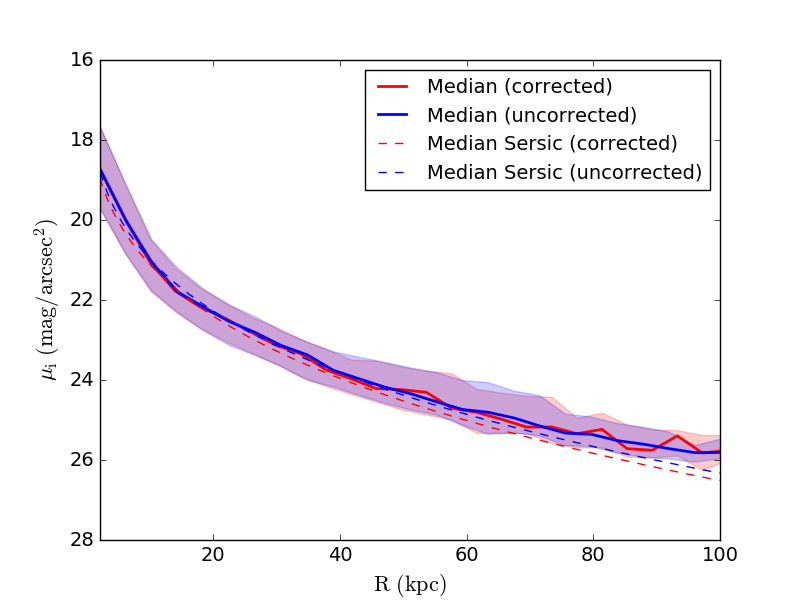}
    \caption{Comparison between the divot corrected and uncorrected stacks as measured by IRAF ellipse. The shaded region represents the $16^{\mathrm{th}}$ and $84^{\mathrm{th}}$ percentiles of the stacks, and the solid lines are the respective medians. The dashed lines are the median S\'{e}rsic model from SIGMA in each respective case. Although not very pronounced here, the S\'{e}rsic models appear to miss some flux on the outskirts of the BCGs, which previous authors have argued is a plateau of either ICL, or a cD halo. There is little difference in the median $n$ values, with values of 4.65 and 4.57 for the non-corrected and corrected models respectively.}    \label{fig:stacky}
\end{figure}
\par As a sanity check, to measure the BCG+ICL profile shape, we fit elliptical isophotes using the \textsc{IRAF} \textsc{ellipse} package (\citealt{1987MNRAS.226..747J}) centred on each cluster BCG, for both the pre- and post- divot-corrected images. The frames are masked at $\mu_{B,rest} = 24 \ \mathrm{mag/arcsec^{2}}$ using the segmentation maps from \textsc{SExtractor} (plus all star/bad pixel masks), due to the convenience of the software having an inbuilt de-blending algorithm to separate object fluxes (with the exception of the BCG itself, which is left unmasked during this process). A stack of the resulting profiles is shown in Figure \ref{fig:stacky}. Interestingly, we do not find much deviation in shape on average when applying a divot correction within the percentiles of the stacks, which supports the comparable outputs we obtained through our \textsc{SIGMA} models.
\par Our results illustrate that one must consider their data carefully when attempting to measure ICL. Indeed, many authors have recognised this issue and have attempted to overcome it by using novel processing methods of their own, such as implementing less `aggressive' global background subtraction techniques (often, for example, using a larger mesh, e.g. \citealt{2018PASJ...70S...6H}, or, a mean global `step', e.g. \citealt{2019MNRAS.482.2838M}). 
\par We recognise that there are several obvious caveats with our method; as aforementioned, surface-brightness methods of measuring ICL tend to recover less flux than methods more readily available in simulations such as setting a binding energy threshold (e.g. \citealt{2011ApJ...732...48R}). Finally, we assume the location of the BCG to be a proxy for the centre of the cluster when measuring the ICL; for local systems, this is often the case (e.g. \citealt{2004ApJ...617..879L}), however the picture has been known to change at high redshift, with higher number of clusters out of dynamical relaxation at $z > 1$ (e.g. \citealt{2011MNRAS.410.1537H}). Having outlined these caveats, we proceed with the view that we have utilised a method which relies as little as possible on parametric modelling; we refer the reader towards arguments for our approach in Section \ref{quant_icl} of this paper.
\begin{table*}
\centering
\caption{A summary of the results from this work, where $f_{ICL}/f_{tot}$ is the percentage of cluster light that is ICL at ${\mu}_{B,rest} = 25 \ \mathrm{mag/arcsec^{2}}$ within $R_{X,500}$ and $\Delta f$ is the fractional difference in the ICL contribution between the divot corrected and uncorrected cases. The equivalent BCG flux ($f_{BCG}$) is also included (S\'{e}rsic model, de Vaucouleurs model and 50 kpc aperture respectively).}
\label{tab:icl_amounts}
\begin{tabular}{l|l|l|l|l|l}
XCS ID                    & $f_{ICL}/f_{tot}$                   & $\Delta f$                        & \begin{tabular}[c]{@{}l@{}}$f_{BCG}$\\ (S\'{e}rsic)\end{tabular} & \begin{tabular}[c]{@{}l@{}}$f_{BCG}$\\ (de Vaucouleurs)\end{tabular} & \begin{tabular}[c]{@{}l@{}}$f_{BCG}$\\ (50 kpc aperture)\end{tabular} \\ \hline
XMMXCS J022456.1-050802.0 & $0.2896   \pm 0.0009$  & $0.0473 \pm 0.0042$  & $0.2829   \pm 0.0007$                                                & $0.2889  \pm 0.0007$                                & $0.2653  \pm  0.0013$                                    \\
XMMXCS J161039.2+540604.0 & $0.1877   \pm 0.0099$  & $0.0547  \pm 0.0091$ & $0.1631  \pm 0.0033$                                                 & $0.1704  \pm 0.0030$                                & $0.1382  \pm  0.0010$                                    \\
XMMXCS J233137.8+000735.0 & $0.2628  \pm 0.0010$    & $0.0626  \pm 0.0076$ & $0.1731   \pm 0.0009$                                                & $0.1568  \pm 0.0006$                                & $0.2469  \pm  0.0025$                                    \\
XMMXCS J232923.6-004854.7 & $0.2757  \pm 0.0009$   & $0.0495 \pm 0.0055$  & $0.1560  \pm 0.0011$                                                 & $0.1657  \pm 0.0010$                                & $0.1201 \pm  0.0008$                                     \\
XMMXCS J161134.1+541640.5 & $0.1540  \pm 0.0006$   & $0.0245 \pm 0.0060$  & $0.0901  \pm 0.0004$                                                 & $0.0871 \pm 0.0003$                                & $0.0690  \pm  0.0007$                                    \\
XMMXCS J095902.7+025544.9 & $0.2676  \pm 0.0012$   & $0.0780  \pm 0.0076$ & $0.1178  \pm 0.0006$                                                 & $0.1159 \pm 0.0005$                                & $0.0895  \pm  0.0008$                                    \\
XMMXCS J095901.2+024740.4 & $0.1148  \pm 0.0017$   & $0.0441 \pm 0.0294$  & $0.3442  \pm 0.0159$                                                 & $0.1705  \pm 0.0012$                                & $0.2058  \pm  0.0028$                                    \\
XMMXCS J100141.6+022538.8 & $0.3121  \pm 0.0007$   & $0.0716  \pm 0.0036$ & $0.2563  \pm 0.0005$                                                 & $0.2535  \pm 0.0005$                                & $0.2254  \pm  0.0013$                                    \\
XMMXCS J095737.1+023428.9 & $0.1567   \pm 0.0007$  & $0.0586  \pm 0.0087$ & $0.1244 \pm 0.0010$                                                  & $0.1291  \pm 0.0009$                                & $0.1535  \pm  0.0012$                                    \\
XMMXCS J022156.8-054521.9 & $0.2887  \pm 0.0012$   & $0.0652  \pm 0.0071$ & $0.2261  \pm 0.0026$                                                & $0.1550  \pm 0.0008$                                & $0.1269  \pm  0.0010$                                    \\
XMMXCS J022148.1-034608.0 & $0.0972 \pm 0.0008$    & $0.0354  \pm 0.0170$ & $0.0670  \pm 0.0012$                                                 & $0.0700 \pm 0.0010$                                & $0.0561 \pm  0.0004$                                     \\
XMMXCS J022530.8-041421.1 & $0.3843  \pm 0.0008$   & $0.0335 \pm 0.0032$  & $0.1660   \pm 0.0007$                                                & $0.1275 \pm 0.0003$                                & $0.1506  \pm  0.0009$                                    \\
XMMXCS J100047.3+013927.8 & $0.2385  \pm 0.0006$   & $0.0391  \pm 0.0041$ & $0.0859  \pm 0.0004$                                                 & $0.0852  \pm 0.0003$                                & $0.0850 \pm  0.0007$                                     \\
XMMXCS J022726.5-043207.1 & $0.2971    \pm 0.0009$ & $0.0337 \pm 0.0048$  & $0.0551 \pm 0.0002$                                                  & $0.0599 \pm 0.0002$                                & $0.0869  \pm  0.0006$                                    \\
XMMXCS J022524.8-044043.4 & $0.3276  \pm 0.0012$   & $0.0627 \pm 0.0059$  & $0.1302  \pm 0.0008$                                                 & $0.1364  \pm 0.0006$                                & $0.0977 \pm  0.0008$                                     \\
XMMXCS J095951.2+014045.8 & $0.1985  \pm 0.0012$   & $0.0410 \pm 0.0100$  & $0.2792  \pm 0.0018$                                                 & $0.2839  \pm 0.0013$                                & $0.1895  \pm  0.0016$                                    \\
XMMXCS J022401.9-050528.4 & $0.2762  \pm 0.0024$   & $0.0334  \pm 0.0131$ & $0.1860  \pm 0.0014$                                                 & $0.1719  \pm 0.0007$                                & $0.1503  \pm  0.0015$                                    \\
XMMXCS J095924.7+014614.1 & $0.3078  \pm 0.0008$   & $0.0542 \pm 0.0042$  & $0.1170  \pm 0.0003$                                                 & $0.1195  \pm 0.0003$                                & $0.1342  \pm  0.0012$  
                            \\ \hline
\textbf{Average} & \textbf{0.2434 $\pm$ 0.0015} & \textbf{0.0475 $\pm$ 0.0085} & \textbf{0.1930 $\pm$ 0.0020} & \textbf{0.1799 $\pm$ 0.0001} & \textbf{0.1642 $\pm$ 0.0013}
\end{tabular}
\end{table*}
\subsection{What Drives ICL Growth?}\label{sec:iclbcg}
\par To enable a more complete interpretation of our results, we perform a partial Spearman analysis on our sample of 18 clusters (see \citealt{2018MNRAS.478.4952F} for method). The partial Spearman enables us to account statistically for underlying correlations which may be present through the means that we have selected our clusters. Here, we choose four primary parameters of interest: the fractional contribution of the ICL and of the BCG ($f_{ICL}/f_{tot}$ and $f_{BCG}/f_{tot}$ respectively), the cluster redshift ($z$) and the cluster mass $M_{X,500}$ (which is computed from the X-ray temperature, as detailed in Section \ref{clusparams_xcs}). We also look at correlations between $k$-corrected BCG absolute magnitude, cluster mass and redshift via a similar means. We hold our significance at the standard value of $p \leq 0.05$ throughout ($\mathrm{log_{10}}[p_{s}] \leq -1.301$). The full Spearman analysis for our clusters is contained in Table \ref{tab:proj2fullspear}; the partial analysis can be found in Appendix B (Tables \ref{t:p2iclfrac}-\ref{t:p2logm500}).
\begin{figure}
    \centering
	\includegraphics[width=\columnwidth]{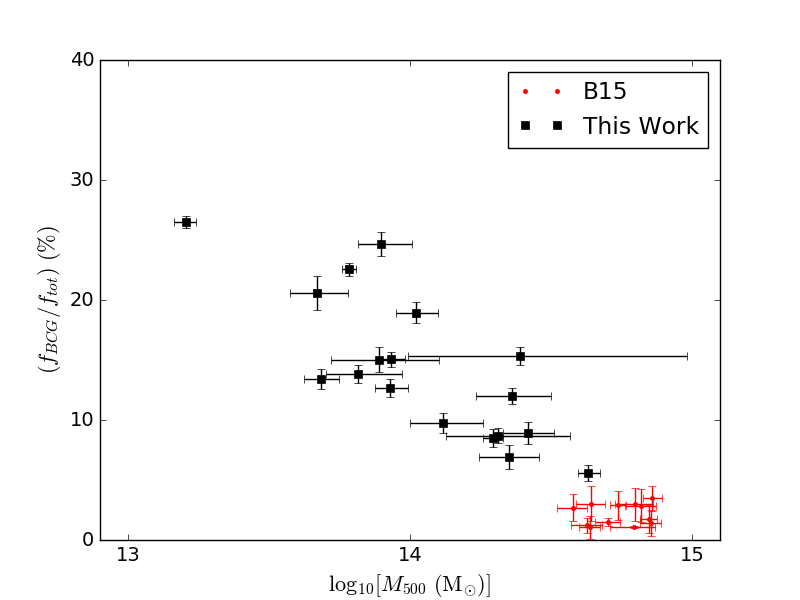}
    \caption{Plot of the BCG to total flux contribution (within a 50 kpc aperture) with respect to halo mass, $M_{X,500}$. The BCG flux contributions from \protect\cite{2015MNRAS.449.2353B} (B15) have been plotted for comparison, which we discuss further in Section \ref{compstudies}. It is clear that there is a strong anti-correlation with halo mass (see text).}
    \label{fig:bcgm500}
\end{figure}
\begin{figure}
    \centering
	\includegraphics[width=\columnwidth]{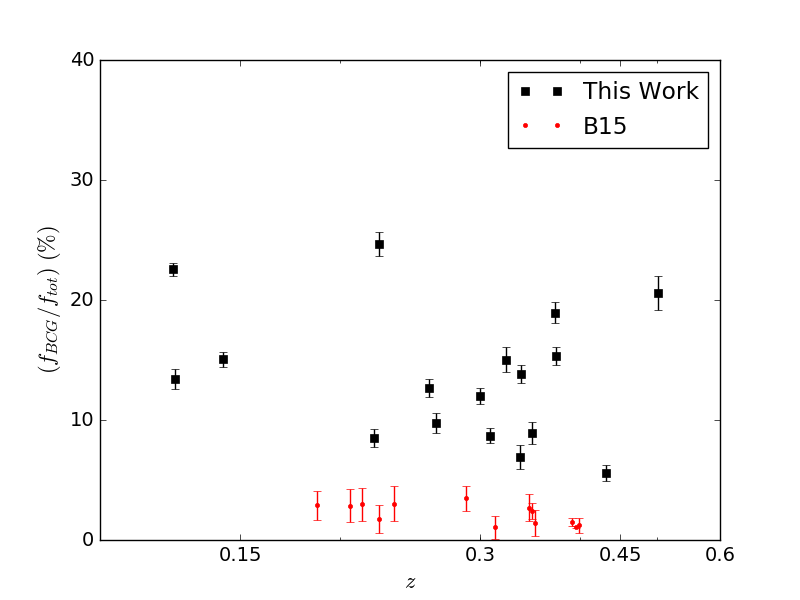}
    \caption{Plot of the BCG to total flux contribution (within a 50 kpc aperture) with respect to redshift, $z$. The BCG flux contributions from \protect\cite{2015MNRAS.449.2353B} (B15) have been plotted for comparison. As shown in the partial Spearman analysis, there is no clear trend with redshift.}
    \label{fig:bcgz}
\end{figure}
\par As aforementioned, in the rest-frame $B$-band, we find a mean ICL flux fraction of around 24\%; this exceeds the mean BCG contribution, even when using a S\'{e}rsic model (16-19\%, see Table \ref{tab:icl_amounts}). Qualitatively however, the difference between the BCG and ICL flux contributions appears to decrease with redshift, with a less than 1\% difference for 2/4 of the most distant systems (with XMMXCS J022148.1-034608.0 being the exception at $\sim$ 4\%) and a reversal of the trend for the highest redshift system at $z = 0.501$. This is not a definitive conclusion, in that we are obviously limited by our small sample size (18 systems) as is the case for most legacy studies of ICL (see references in the Introduction), alongside significant caveats with assuming a fixed aperture scale when measuring the fluxes of our BCGs. However, it raises interesting questions as to what point in time the ICL begins to dominate the cluster halo (see Section \ref{compstudies}).
\begin{table}
\centering
\caption{Full Spearman analysis of all the parameters used in this study: the fractional contribution of the ICL and of the BCG ($f_{ICL}/f_{tot}$ and $f_{BCG}/f_{tot}$ respectively), the cluster redshift ($z$) and the cluster mass $M_{X,500}$. The top half of the table lists the Spearman rank correlation coefficient ($r_{s}$), whereas the bottom half of the table provides the log of its corresponding p-value ($\mathrm{log_{10}}[p_{s}]$, expressed as such due to some p-values being very small).}
\begin{tabular}{lllll}
                             & $f_{ICL}/f_{tot}$ & $f_{BCG}/f_{tot}$ & $z$      & $\mathrm{log_{10}}M_{X,500}$ \\
$f_{ICL}/f_{tot}$            & $-$               & 0.0807          & -0.7860 & -0.2070                     \\
$f_{BCG}/f_{tot}$            & -0.1292          & $-$               & -0.1526 & -0.7474                     \\
$z$                          & -4.0174           & -0.2746          & $-$      & 0.3561                      \\
$\mathrm{log_{10}}M_{X,500}$ & -0.4051           & -3.4491           & -0.8700 & $-$                         
\end{tabular}
\label{tab:proj2fullspear}
\end{table}
\par In common with other authors (e.g. \citealt{2015MNRAS.449.2353B} and upcoming discussion), we detect a significant anti-correlation ($r_{s} = -0.786$, $\mathrm{log_{10}}[p_{s}] = -4.017$) between the contribution of ICL with cluster redshift, which remains almost entirely unchanged when fixing for cluster mass (see Table \ref{t:p2logm500} in the Appendix). This is clearly visible on Figures \ref{fig:icl+bcgz2} and \ref{fig:iclz}, which we will discuss in Section \ref{compstudies}. This is not the case for the BCG flux fraction, which has no significant correlation with redshift ($r_{s} = -0.153$, $\mathrm{log_{10}}[p_{s}] = -0.275$, see Figure \ref{fig:bcgz}) and remains highly anti-correlated with the cluster mass even after fixing for redshift ($r_{s} = -0.750$, $\mathrm{log_{10}}[p_{s}] = -3.477$, see Table \ref{t:p2z} in the Appendix and Figure \ref{fig:bcgm500}). Even if we consider a S\'{e}rsic model (which produces almost universally the largest BCG fraction estimates) in place of an aperture magnitude for our BCGs, there is still an anti-correlation present at fixed redshift that remains almost unchanged ($r_{s} = -0.775$, $\mathrm{log_{10}}[p_{s}] = -3.801$), so the trend is robust to the flux loss through not accounting for galaxy profile wings. 
\begin{figure*}
    \centering
	\includegraphics[width=16cm,height=16cm,keepaspectratio,,trim={1cm 2cm 1cm 2cm},clip]{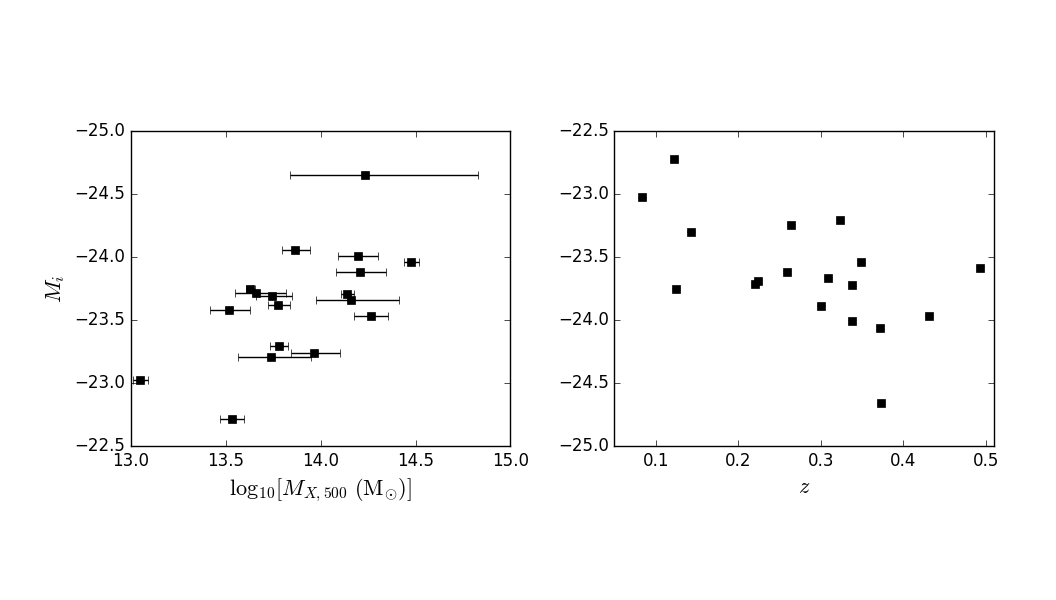}
    \caption{The rest-frame BCG absolute magnitude ($M_{i}$, $i$-band aperture) versus cluster mass measured in X-rays ($M_{X,500}$, left) and redshift ($z$, left). For ease of comprehension, we have inverted the $y$-axis. No significant trends are detected in our partial Spearman analysis between BCG magnitude with either redshift of halo mass.}
    \label{bcgabsmag_mx500z}
\end{figure*}
\par There is no strong correlation present, however, between the ICL and the mass of the cluster at fixed redshift ($r_{s} = 0.126$, $\mathrm{log_{10}}[p_{s}] = $ from Table \ref{t:p2z}). This has an interesting implication, in that our findings imply a much closer dependence between stars within the very central region of the halo (BCG) with the halo properties (such as $M_{500}$) in comparison to stellar mass distributed further out (ICL). Indeed, with a lack of correlation present between halo mass and ICL mass, there seems to be a `decoupling' between the two components; the ICL, for instance, has been found to exhibit far more growth since $z \sim 1$ than the BCG (e.g. \citealt{2012MNRAS.425.2058B, 2015MNRAS.449.2353B}), with BCG growth rates being much more modest than those predicted from simulations. 
\par The leftmost panel of Figure \ref{bcgabsmag_mx500z} shows the relationship between the $k$-corrected BCG absolute magnitude ($M_{i}$, $i$-band aperture, see Section \ref{compstudies}) and cluster mass ($M_{X,500}$). Although we detect an anti-correlation between halo mass and absolute magnitude (which is anti-correlated with BCG mass), it is not significant ($r_{s} = -0.40877$, $\mathrm{log_{10}}[p_{s}] = -1.0785$). This finding is also the case if we fix for redshift ($r_{s} = -0.27456$, $\mathrm{log_{10}}[p_{s}] = -0.56831$). If we remove the two points with the largest errorbars, it becomes significant by our criteria, but still remains insignificant with fixed redshift ($r_{s} = -0.48775$, $\mathrm{log_{10}}[p_{s}] = -1.3094$; $r_{s} = -0.37136$, $\mathrm{log_{10}}[p_{s}] = -0.80487$ respectively). We therefore do not find conclusive evidence that our BCG absolute magnitudes (and therefore masses) are strongly governed by halo mass here. This is likely to be as a result of our selection (e.g. \citealt{2015MNRAS.449.2353B}) and also due to the fact that our sample size is small. An obvious point would therefore be to establish whether our result for the BCG flux fraction with halo mass weakens when applying our method a larger sample of clusters with an established $M_{\mathrm{BCG}}-M_{halo}$ relation; this was also recognised in \cite{2015MNRAS.449.2353B}.
\par We find a similar result for absolute magnitude with redshift when fixing for halo mass ($r_{s} = -0.46034$, $\mathrm{log_{10}}[p_{s}] = -1.2632$, see rightmost panel of Figure \ref{bcgabsmag_mx500z}) even having removed the two points with the largest error bars ($r_{s} = -0.31443$, $\mathrm{log_{10}}[p_{s}] = -0.62785$); hence, we do not detect any significant change in BCG brightness with redshift either. Although this may also be linked to the way we have selected our BCGs, given numerous authors have found little change in BCG brightnesses since $z \sim 1$ (e.g. \citealt{2008MNRAS.387.1253W}, \citealt{2009Natur.458..603C}, \citealt{2010ApJ...718...23S}), our result acts to support trends found by other works using independent data sets.
\subsection{Comparison with Other Studies}\label{compstudies}
\par
\begin{figure}
    \centering
	\includegraphics[width=\columnwidth]{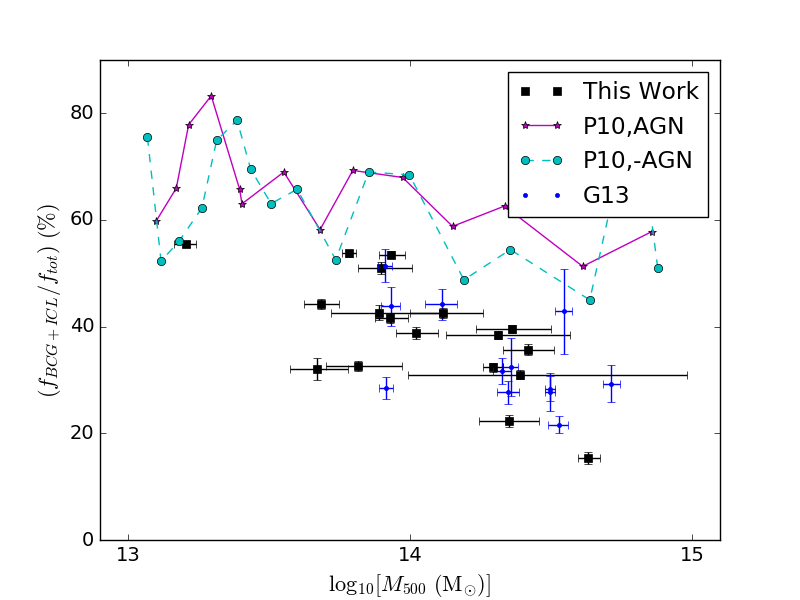}
    \caption{Comparison of the relative fluxes of ICL+BCG versus halo mass. The legend key is as follows: \protect\cite{2010MNRAS.406..936P} (simulation) with/without an AGN prescription (P10, AGN/-AGN) and \protect\cite{2013ApJ...778...14G} (G13, observational). It is clear that P10 does not agree with either our observational results or the results of G13.}
    \label{fig:icl+bcgm500}
\end{figure}

\begin{figure}
    \centering
	\includegraphics[width=\columnwidth]{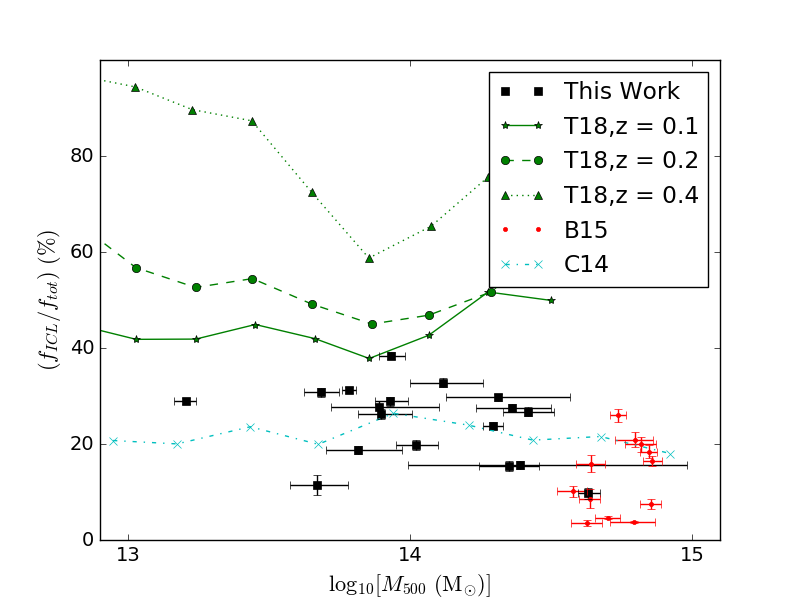}
    \caption{As in Figure \ref{fig:icl+bcgm500}, but for ICL flux only (see text). The legend key is as follows: \protect\cite{2018ApJ...859...85T} at redshift $z$ (T18, simulation), \protect\cite{2015MNRAS.449.2353B} (B15, observational) and \protect\cite{2014MNRAS.437.3787C} (C14, simulation). With the exception of C14, there is a clear disagreement between the observations and simulations.}
    \label{fig:iclm500}
\end{figure}

\begin{figure}
    \centering
	\includegraphics[width=\columnwidth]{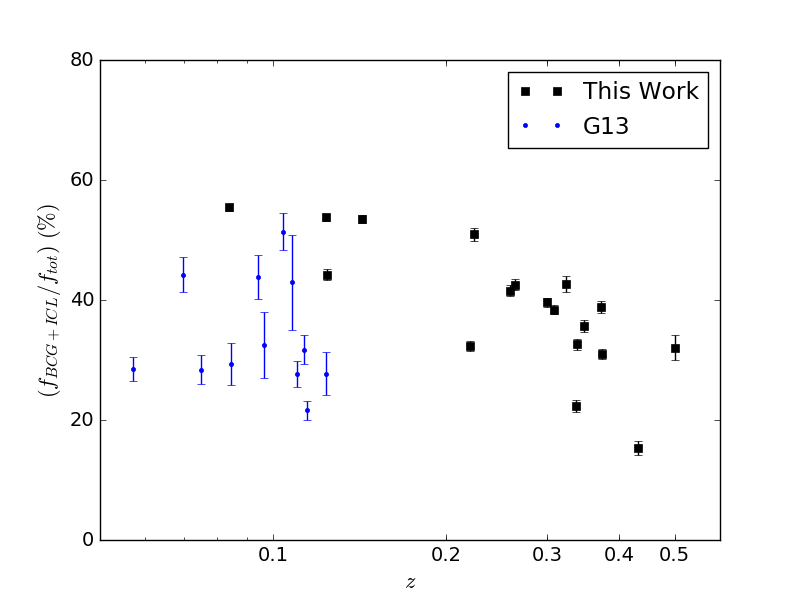}
    \caption{Comparison of the relative fluxes of ICL+BCG versus redshift, with the points from \protect\cite{2013ApJ...778...14G} (G13).}
    \label{fig:icl+bcgz2}
\end{figure}

\begin{figure}
    \centering
	\includegraphics[width=\columnwidth]{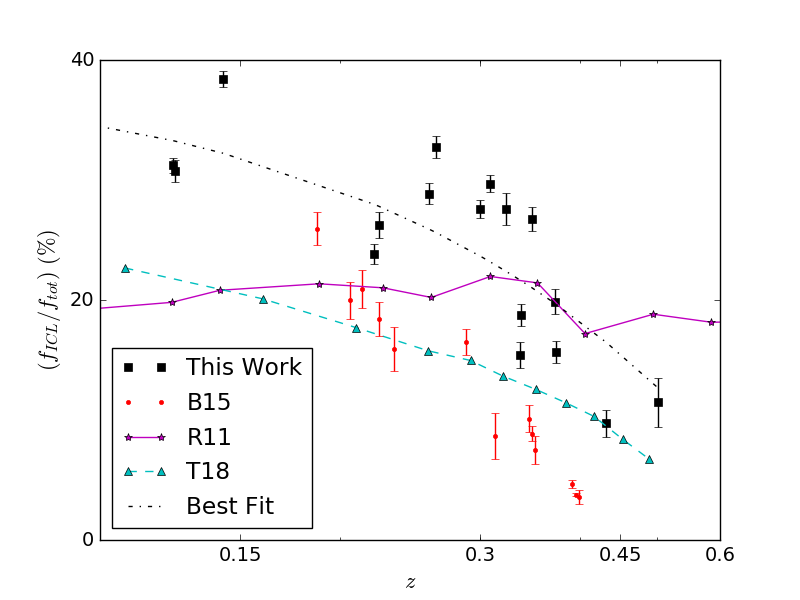}
    \caption{As in Figure \ref{fig:icl+bcgz2}, but for ICL flux only. The legend key is as follows: \protect\cite{2015MNRAS.449.2353B} (B15), \protect\cite{2011ApJ...732...48R} (R11) and \protect\cite{2018ApJ...859...85T} (T18). The best least-squares fit has been included for comparison (slope = -54.50, intercept = +40.01).}
    \label{fig:iclz}
\end{figure}

 We show the results from a number of other studies of ICL, from both simulations and observations, in Figures \ref{fig:icl+bcgm500}$-$\ref{fig:iclz} alongside our results. Where relevant, we have included descriptions giving context to the results presented in the plots. The shorthand for the observational studies shown in the legends of the plots is: \cite{2013ApJ...778...14G} (G13, parametric model) and \cite{2015MNRAS.449.2353B} (B15, ${\mu}_{B} = 25 \ \mathrm{mag/arcsec^{2}}$). Respectively, the shorthand for the simulation-based studies presented in the legends of the plots is: \cite{2010MNRAS.406..936P} (P10, both with and without an AGN feedback prescription applied), \cite{2011ApJ...732...48R} (R11, ${\mu}_{V} = 25 \ \mathrm{mag/arcsec^{2}}$), \cite{2014MNRAS.437.3787C} (C14, disruption model only) and \cite{2018ApJ...859...85T} (T18, ${\mu}_{V}=24.7 \ \mathrm{mag/arcsec^{2}}$, mock SDSS $r$-band; closest to our own data). All observational masses have been scaled from X-ray measurements (from either XMM Newton or Chandra in the case of the majority of the CLASH clusters) using the same scaling relation (\citealt{2005A&A...441..893A}). In the case of the CLASH sample, it is worth noting that clusters with $T>5$ keV have a $\sim$15\% mass increase on average between values computed from Chandra vs XMM-Newton data (see discussion in \citealt{2018MNRAS.474.3009D} and \citealt{2013ApJ...767..116M}), however, scaling the points does little to influence the interpretation of our comparisons (see upcoming discussion). In the case of the theoretical studies, the density contrast was scaled where necessary (e.g. from ${\rho}_{c}=200$ to ${\rho}_{c}=500$) using the method outlined in \cite{2003ApJ...584..702H}, assuming an NFW profile with a concentration of 3. 
\par Figures \ref{fig:icl+bcgm500} and \ref{fig:iclm500} show the relationship between the BCG+ICL fraction and the ICL fraction with cluster mass respectively. In both cases, there is an obvious difference between the results from simulations and observations, in that while the observations qualitatively appear fairly consistent (see upcoming discussion) the simulations appear to predict significantly larger BCG+ICL (or ICL) contributions to the overall cluster light. The exception here is \cite{2014MNRAS.437.3787C}, whose results are consistent with observations (Figure \ref{fig:iclm500}); their simulations are however semi-analytic rather than hydrodynamic. Although not plotted here, larger BCG+ICL fractions than those seen observationally (60-80\% compared with 1-60\%) were also found by \cite{2014MNRAS.437..816C}, who, using hydrodynamical simulations (for specifics, see \citealt{2011MNRAS.418.2234B}), measured the BCG+ICL light using a $V$-band surface brightness limit, similar to our own approach. \cite{2014MNRAS.437.3787C} also found that their ICL fractions were also very sensitive to AGN and supernova feedback; this is in contrast to \cite{2010MNRAS.406..936P}, as while the BCG+ICL fractions themselves are similar, there was little difference found between the fractions detected when an AGN model was used (see Figure \ref{fig:icl+bcgm500}). An exception using simulations is \cite{2007MNRAS.377....2M}, who found a much lower average fraction of ICL with respect to halo mass ($\sim$ 22\%) in a similar mass range to that of the CLASH clusters ($10^{14}-10^{15} \ \mathrm{M_{\odot}}$); however, they found a positive correlation between halo mass and the fraction of ICL, which has not been seen observationally. In fact, the opposite has increasingly been reported, with lower mass haloes found to be more `efficient' producers of stellar mass than large clusters (e.g. Figure 8 of \citealt{2018ApJ...859...85T}, simulations; \citealt{2018MNRAS.474.3009D} and  \citealt{2019A&A...631A.175E}, both observations).
\par Results for the ICL light fraction with respect to the overall cluster from numerical simulations and SAMs appear generally to be more self-consistent than those obtained observationally (e.g. for our work, 20-40\%, see \citealt{2014MNRAS.437.3787C}, \citealt{2011ApJ...732...48R} for some typical SAM results). \cite{2009JApA...30....1B}, using a numerical prescription, simulated the build-up of intracluster stars using several different cluster mass profiles (e.g. Perseus-like to Virgo-like) while considering the morphology of the galaxies contained within the cluster (e.g. if the BCG was a cD-type). They found mean ICL fractions of $\sim$ 25\% for a Virgo- or Perseus-like system, compared with much higher fractions ($\sim$ 40\%) for an NFW model; they also found a dependence of the ICL fraction on the morphology of the BCG (with cD-type BCGs leading to generally more centrally-concentrated ICL profiles). \cite{2010MNRAS.403..768H}, building on the semi-analytic study of BCG mass growth of \cite{2007MNRAS.375....2D}, included a prescription for ICL (tidal disruption and dynamical friction); they found a mean fraction of ICL of around 18\%, with a positive correlation between the ICL fraction and the halo mass of the cluster. \cite{2014MNRAS.437.3787C} used dark matter haloes from the Millennium Simulation, coupled with several simple dynamical models (e.g. mergers, disruption, tidal stripping), finding results similar to observations in \cite{2013ApJ...778...14G} (and indeed, our own), with no correlation with halo mass.
\par Here, we detect no strong trend between halo mass and the fraction of ICL (as is the case in \citealt{2015MNRAS.449.2353B}); it is therefore possible that any gradients present in Figure \ref{fig:icl+bcgm500} are driven by the strong anti-correlation between the BCG flux fraction and halo mass established in Section \ref{resultsiv}, as they are not present with the ICL fraction itself. \cite{2011MNRAS.414..602T} also find little evidence dynamically for any strong relation between the BCG+ICL fraction with cluster mass for a single cluster at $z \sim 0.3$, acquiring a total fraction of $\sim 70$\% in line with the low-$z$ ($z < 0.1$) results of \cite{2007ApJ...666..147G}, though higher than the latter's 2013 revisited study (\citealt{2013ApJ...778...14G}, see Figures \ref{fig:icl+bcgm500}-\ref{fig:icl+bcgz2}) and indeed, our own work. Our results with respect to halo mass are fairly consistent observationally with the other studies presented in Figures \ref{fig:icl+bcgm500} and \ref{fig:iclm500}, with the $\sim$ 24\% ICL fraction and $\sim$ 41\% BCG+ICL fraction seen here comparable with the respective results of \cite{2015MNRAS.449.2353B} and \cite{2013ApJ...778...14G}. As discussed at length however in the Introduction, observational results for ICL dramatically vary in general, with a dependence on the data used and the measurement approach (see Figure 8 of \citealt{2018ApJ...859...85T}). 
\par \cite{2018ApJ...859...85T} investigated the limitations of measuring ICL from optical imaging data using hydrodynamical simulations. Although their ICL result differs significantly from our own and that of \cite{2015MNRAS.449.2353B}, their findings on the causes of what effects drive scatter in ICL measurements are arguably far more interesting. Using simulated images of their clusters, they produced mock images with numerous observational differences, such as band, pixel size, surface brightness limit and PSF size. They found a clear effect from the PSF, finding that large PSFs lead to greater smoothing and a slightly higher ICL fraction (5-10\%, see upcoming discussion). They also found a band dependence on the ICL fraction, finding that the $r$-band yielded a much larger ICL fraction ($\sim 2 \times$) even when using the same equivalent $V$-band surface brightness limit; they attribute this in their discussion to uncertainties in their stellar population model of choice (\citealt{2003MNRAS.344.1000B} with a \citealt{2003PASP..115..763C} IMF). They also found that the surface brightness limit also affected their ICL result, finding a doubling in the amount of ICL detected between $23.0 < {\mu}_V < 26.5$ for low redshift haloes (also observed in \citealt{2014MNRAS.437..816C}, from whom their method for generating mock images was derived). Finally \cite{2018ApJ...859...85T} also found a clear dependence of cosmological dimming on their ICL, finding an $increase$ in the relative fraction of ICL up to $z \sim 1$ when accounting for surface brightness dimming (see rightmost panel of their Figure 6). Their results suggest a clear motivation for more studies of this kind, as such a result has unexpected consequences regarding the current widely-accepted paradigm of BCG-ICL co-evolution (see Introduction), given it is canonically thought that the period $0 < z < 1$ is an era of rapid ICL growth, with little changes in the luminosity of the BCG.
\par The theoretical studies presented here also obviously differ enormously in their methodology, with some using methods to estimate ICL that are not observationally feasible (such as tracking star particles). It is, however, curious that despite more complex physical models being included in hydrodynamical simulations, they generally seem to struggle to reproduce ICL fractions with cluster mass in contrast to either a simple numerical or semi-analytic prescription. This therefore presents a challenge to these modern simulation suites and an opportunity for further analysis to better understand the reasons behind these differences, such as the effects of subgrid size and the physical models used. Future studies resembling this work with larger cluster samples (e.g. in the wake of the Vera C. Rubin Observatory) will also help in our understanding of these discrepancies.  
\par Figures \ref{fig:icl+bcgz2} and \ref{fig:iclz} show the trend of our results with redshift, for the BCG+ICL fraction and ICL fraction respectively. Although we appear consistent with \cite{2013ApJ...778...14G} in Figure \ref{fig:icl+bcgz2}, there is some deviation present between our results and those presented in Figure \ref{fig:iclz} (e.g. \citealt{2015MNRAS.449.2353B}), in that our fractions with redshift appear noticeably higher. Interestingly, however, there seems to be no clear consensus overall, with the slopes of ICL growth differing clearly across studies. There may be several reasons as to why this may be the case. Firstly, as noted in \cite{2018ApJ...859...85T}, observational results are strongly influenced by several factors. The PSF, for example, was found by \cite{2018ApJ...859...85T} to produce a scatter of 5-10\% in the total ICL fraction at a redshift range similar to that explored here ($0 < z < 0.4$ from their Figure 3, ${\mu}_V = 26.5$), with a smaller PSF (such as those found using space-based observatories as in CLASH) and usually also produced smaller results for the ICL fraction. They also found that measuring the ICL in a redder passband (SDSS $r$) increased the fraction of ICL detected, even when using the same equivalent threshold, by around a factor of 2.
\par The results from \cite{2018ApJ...859...85T} presented in Figure \ref{fig:iclz} represent an SDSS-like, $V$-band image with the ICL measured using an isophote of ${\mu}_{V} = 24 \ \mathrm{mag/arcsec^{2}}$, with cosmological surface brightness dimming not being taken into account. This corresponds to a growth factor of $\sim$ 3, similar to what we observe. As previously mentioned, \cite{2018ApJ...859...85T} do not actually find any physical growth of the ICL over cosmic time; in fact, they find the ICL contribution to dramatically shrink with decreasing redshift. This result sets them starkly apart from most other theoretical studies, which, within the redshift range explored in this work, find fractional increases in the ICL relatively consistent with our own of 1.5-4 (e.g. \citealt{2004MNRAS.355..159W}; \citealt{2007MNRAS.377....2M}; \citealt{2011ApJ...732...48R}; \citealt{2014MNRAS.437.3787C}).
\par Some of the observational reasons outlined in \cite{2018ApJ...859...85T} may partially account for the difference we see between our results and the results of \cite{2015MNRAS.449.2353B}. We have, for example, larger $k$-corrections due to our use of a redder band (HSC-$i$); testing EZGAL using the bands used in CLASH (F606W, F626, F775, F850LP) with an identical stellar evolution model however produced similar results to \cite{2015MNRAS.449.2353B}, with the same trends. Our data is also ground-based with a larger PSF ($<\mathrm{FWHM}> \sim 0.56 ^{\prime \prime}$ for HSC, see \citealt{2018PASJ...70S...8A}), which we correct for when fitting profiles (but we do not deconvolve our data when computing ICL); the effect of this on recovering the magnitudes in HSC-SSP-Wide data was investigated in detail by \cite{2018PASJ...70S...6H}, where they determined a 10-18\% margin of error in $i$-band magnitudes at $25^{\mathrm{th}}$ mag (see also \citealt{2018MNRAS.475.3348H}). HSC-SSP-Deep is $\sim$ 1 magnitude deeper than that of CLASH (where, as noted in \citealt{2015MNRAS.449.2353B}, a difference in 0.5 magnitudes in survey depth results in a 5-10\% reduction in the amount of measured ICL component). Our values are, of course, also divot-corrected. 
\par One of the biggest differences we observe is that of the fractional contribution of the BCG, in that ours are far larger than those stated in \cite{2015MNRAS.449.2353B} ($\sim$ 19\% compared to $\sim$ 5\%, see Figure \ref{bcgabsmag_mx500z}). It is not clear from \cite{2015MNRAS.449.2353B} whether the fractions are measured relative to a set absolute cluster radius (here, $R_{X,500}$ in kpc); however, using a radius of $R_{500}$, similar low BCG fractions are seen in \cite{2012MNRAS.425.2058B} (2-4\% depending on whether a de Vaucouleurs model or 50 kpc aperture is used). This differs significantly from numerous other works, with which our BCG fractional contribution to the overall cluster light is more consistent; it is, for example, comparable to \cite{2005MNRAS.358..949Z} at $z \sim 0.25$, who fit de Vaucouleurs profiles and measure the relative fractions contained within a fixed radius of 500 kpc centred on the BCG. A BCG stellar mass contribution to the overall halo of around 15-40\% was also noted by \cite{2015ApJ...800..122S}, as well as in \cite{2007MNRAS.378.1575S}. The CLASH sample constitutes especially massive systems, with the range of cluster masses representing the larger end of the cluster population ($\sim 10^{15} \ M_{\odot}$); no overlap is present with our sample. As other authors have shown (e.g. \citealt{2010MNRAS.407..263A}, \citealt{2019A&A...631A.175E}), there appears to be an increasing inefficiency in stellar mass production with increasing halo mass, particularly with respect to the BCG.
\par There is also the added issue of how CLASH data was optimised for science, in that its original focus was specifically to study the lensed and high-$z$ Universe, rather than LSB science (\citealt{2012ApJS..199...25P}). The background subtraction method is therefore generally more aggressive than ideal (although there is a focus on lensing in HSC, there is also an LSB science focus and a great deal of pipeline refinement in preparation for the Vera C. Rubin Observatory). In addition, HST's ACS has a far smaller field of view ($\sim 3.36^{\prime}$) than HSC ($\sim 1.5^{\circ}$), as well as a very small associated dither pattern in CLASH. The majority of the clusters within CLASH (given their redshift) therefore would fill the majority of a frame (for example, a cluster with a radius of 0.7 Mpc at $z = 0.4$ has an angular extent of $2.2^{\prime}$, assuming our concordance cosmology). This implies that it is unlikely that the true background is reached (i.e. that there is little available sky with respect to source), leading to an overestimate of the background. As a further example, the `missing flux' issue with HST WFC3 (which has a smaller FOV than the ACS at $2.7^{\prime}$) data was explored in detail in \cite{2019A&A...621A.133B}, who produced a pipeline to re-reduce Hubble Ultra-Deep Field data (\citealt{2006AJ....132.1729B}); they found, when re-reduced, an integrated magnitude of recovered light of $\sim 20$ mag, which they state is comparable to the brightest galaxies in the field. Although \cite{2019A&A...621A.133B} did not apply their method to the CLASH data, it is likely, given the comparable observational and data-reduction methodology, that it suffers from the same issue. This issue in particular may well be a large factor in the difference between the results of \cite{2015MNRAS.449.2353B} and our own. With its larger FOV, HSC SSP DR1 is more appropriate for LSB science (see, for example, Figure 5 of \citealt{2018PASJ...70S...8A}); moreover, further pipeline processing improvements have been made with the release of DR2 (see, for example, discussions in \citealt{Aihara_2019}). Work is currently underway to establish how effective the DR2 pipeline will be for the deeper data stream from the Vera C. Rubin Observatory (e.g. Watkins et al., in prep.).
\section{Summary and Conclusions}
\par In this work, we measured the ICL in 18 XCS-HSC clusters alongside consideration of two systematics: background contribution and sky over-subtraction. We discussed the sample of clusters in XCS used in this work and how they were selected; we also discussed the HSC-SSP survey, the current processing pipeline and the photometry used. We outlined how we measured our ICL, using an equivalent $B$-band isophotal threshold, measured within an aperture of radius $R_{X,500}$ centred on the brightest cluster galaxy. We introduced the `divot' problem, which arises due to an `over-subtraction' of flux from background estimation during image processing, and our method to correct for this effect. Finally, we introduced a set of basic simulations to allow us to understand the flux contribution from the background at a given mean surface brightness for an ICL-like profile. 
\par We then presented our results alongside numerous other studies for comparison, from simulations and observations. We noted a large degree of scatter, observationally (1-60\% globally, 20-40\% for our sample) and theoretically (10-90\%) for retrieved ICL fractions. We then discussed at length some of the reasons as to why such discrepancies may exist, such as the data used, the measurement methodology, the simulation method and so on. Our primary conclusions are as follows:
\begin{enumerate}
    \item there is a loss in ICL flux of about $4 \times$ the estimated background from the effects of sky over-subtraction, which remains approximately constant $\pm 1 \mathrm{mag/arcsec^{2}}$ about our lowest chosen threshold. We surmised that this was likely to be an underestimate, given the S\'{e}rsic models used when creating the divot corrections, 
    \item the divot corrections themselves do little to change the overall profile shape, with the 1-D profiles and parametric fits from \textsc{SIGMA} respectively yielding very similar results,
    \item we detect no significant correlation between BCG absolute magnitude and redshift when fixing for halo mass. We also find the fractional contribution of BCG light with respect to the overall cluster light to be strongly anti-correlated with halo mass, inferring that star-formation efficiency is inversely proportional to halo mass (e.g. \citealt{2019A&A...631A.175E}), 
    \item we find no strong evidence that the contribution of ICL to the overall stellar content of the cluster is strongly linked to halo mass, in line with most recent simulations, 
    \item the fraction of ICL light is not strongly linked to the fractional contribution of the BCG (Section \ref{resultsiv}), indicating a `decoupling' between the two components (e.g. \citealt{2018MNRAS.474.3009D}), 
    \item while finding generally higher fractions with redshift, we find the ICL to grow by a factor of $\sim$ 2-4 between $0.1 < z < 0.5$, slightly more modest than the factor of 4-5 in clusters over a similar range in redshift from \cite{2015MNRAS.449.2353B} albeit with a higher scatter ($rms \ f_{ICL}/f_{tot} \sim$ 10\%),
    \item we find a significant difference generally between hydrodynamical stellar mass fractions of ICL and BCG+ICL in clusters at a given halo mass, with the simulations almost always over-predicting the contribution (even when measured in an observationally consistent way). Numerical and SAM based simulations, however, yield results closer to our observations. 
\end{enumerate}
\par Our work supports the current scenario of relatively rapid ICL build-up since $z \sim 0.5$, with BCGs remaining relatively unchanged with respect to absolute magnitude. There is also, as has been the case for most observational studies, discrepancies present between simulations. From the evidence presented here, it seems that a far greater understanding of the observational effects involved is needed (e.g. surface brightness limit used, band used, whether a BCG+ICL model fit is used and PSF size), given that such effects, as noted in \cite{2018ApJ...859...85T}, can change the ICL result obtained by a factor of 2. 
\par As sample sizes grow larger and publically-available image data improves in depth with the new generation of telescopes in the coming decade (such as the Vera C. Rubin Observatory, which promises frequent periodic imaging of the whole southern sky coupled with an enormous FOV of $9.62 \ \mathrm{deg^{2}}$, see \citealt{2008arXiv0805.2366I} and \citealt{brough2020vera}), studies will be more readily able to untangle the degeneracies, for example, between a detection of ICL growth and the effect of surface brightness dimming. For now, however, our results support the paradigm of ICL growth being the dominant stellar evolutionary component in galaxy clusters since $z \sim 1$.

\section{Data Availability}
\par The data underlying this article will be shared on reasonable request to the corresponding author.
\section*{Acknowledgements}
We would like to thank the referee for their detailed comments which much improved and clarified this paper. K.E. Furnell would like to thank the XCS team for their cooperation, collaboration and support throughout this work. K.E. Furnell would also like to thank L.S. Kelvin for his consistent advice, support and for enabling the use of the \textsc{SIGMA} software. Finally, K.E. Furnell would like to thank P. Bradshaw for providing constant support, kindness and encouragement throughout her PhD. K.E. Furnell is supported by a Science and Technologies Funding Council (STFC) award. C.A. Collins, I.K. Baldry and L.S. Kelvin are supported by an STFC research grant (ST/M000966/1). PTPV was supported by Funda\c{c}\~{o} para a Ci\^{e}ncia e a Tecnologia (FCT) through research grants UIDB/04434/2020 and UIDP/04434/2020. 



\bibliographystyle{mnras}
\bibliography{bibliography} 
\bsp
\clearpage



\appendix
\section*{Appendix A - Alternative $k_{i,B}$-correction parameters}\label{appendixa}
\begin{table}
\centering
\caption{The mean $k_{i,B}$ correction values for 3 formation redshifts $z_{f}$ at 3 metallicity values (where $Z_{\odot}$ = solar) across all BC03 models for all sample clusters within $0 < z < 0.28$. The mean for the clusters used in this work is in the centre-left cell of this table.}
\begin{tabular}{llll}
            & $Z = Z_{\odot}$ & $Z = 0.4 Z_{\odot}$ & $Z = 2.5 Z_{\odot}$ \\
$z_{f} = 2$ & -1.4409      & -1.3203                          & -1.6470                          \\
$z_{f} = 3$ & -1.4687      & -1.3428                          & -1.6679                          \\
$z_{f} = 4$ & -1.4811     & -1.3531                         & -1.6822                         
\end{tabular}
\label{t:eklowz}
\end{table}
\begin{table}
\centering
\caption{As in Table \ref{t:eklowz}, but for clusters within $0.28 < z < 0.5$.}
\begin{tabular}{llll}
            & $Z = Z_{\odot}$ & $Z = 0.4 Z_{\odot}$ & $Z = 2.5 Z_{\odot}$ \\
$z_{f} = 2$ & -1.2495     & -1.1619         & -1.4092          \\
$z_{f} = 3$ & -1.2735      & -1.1755         & -1.4127          \\
$z_{f} = 4$ & -1.2865      & -1.1849          & -1.4181         
\end{tabular}
\label{t:ekhighz}
\end{table} 

\section*{Appendix B - Partial Spearman Analysis}
\label{section:partspearproj2}
\begin{table}
\centering
\caption{Partial Spearman analysis for the parameters discussed in Section \ref{resultsiv}, with $f_{ICL}/f_{tot}$ held constant.}
\begin{tabular}{lllll}
                             & $f_{ICL}/f_{tot}$ & $f_{BCG}/f_{tot}$ & $z$      & $\mathrm{log_{10}}M_{X,500}$ \\
$f_{ICL}/f_{tot}$            & $-$               & $-$               & $-$      & $-$                          \\
$f_{BCG}/f_{tot}$            & $-$               & $-$               & -0.1448 & -0.7493                     \\
$z$                          & $-$               & -0.2467          & $-$      & 0.3198                      \\
$\mathrm{log_{10}}M_{X,500}$ & $-$               & -3.4625           & -0.7082  & $-$                         
\end{tabular}
\label{t:p2iclfrac}
\end{table}
\begin{table}
\centering
\caption{Partial Spearman analysis for the parameters discussed in Section \ref{resultsiv}, with $f_{BCG}/f_{tot}$ held constant.}
\begin{tabular}{lllll}
                             & $f_{ICL}/f_{tot}$ & $f_{BCG}/f_{tot}$ & $z$      & $\mathrm{log_{10}}M_{X,500}$ \\
$f_{ICL}/f_{tot}$            & $-$               & $-$               & -0.7854 & -0.2215                     \\
$f_{BCG}/f_{tot}$            & $-$               & $-$               & $-$      & $-$                          \\
$z$                          & -3.9488           & $-$               & $-$      & 0.3687                      \\
$\mathrm{log_{10}}M_{X,500}$ & -0.4237          & $-$               & -0.8787  & $-$                         
\end{tabular}
\label{t:p2bcgfrac}
\end{table}
\begin{table}
\centering
\caption{Partial Spearman analysis for the parameters discussed in Section \ref{resultsiv}, with $z$ held constant.}
\begin{tabular}{lllll}
                             & $f_{ICL}/f_{tot}$ & $f_{BCG}/f_{tot}$ & $z$ & $\mathrm{log_{10}}M_{X,500}$ \\
$f_{ICL}/f_{tot}$            & $-$               & -0.0642         & $-$ & 0.1262                      \\
$f_{BCG}/f_{tot}$            & -0.0969          & $-$               & $-$ & -0.7504                     \\
$z$                          & $-$               & $-$               & $-$ & $-$                          \\
$\mathrm{log_{10}}M_{X,500}$ & -0.2091          & -3.4767           & $-$ & $-$                         
\end{tabular}
\label{t:p2z}
\end{table}
\begin{table}
\centering
\caption{Partial Spearman analysis for the parameters discussed in Section \ref{resultsiv}, with $\mathrm{log_{10}}M_{X,500}$ held constant.}
\begin{tabular}{lllll}
                             & $f_{ICL}/f_{tot}$ & $f_{BCG}/f_{tot}$ & $z$      & $\mathrm{log_{10}}M_{X,500}$ \\
$f_{ICL}/f_{tot}$            & $-$               & -0.1139          & -0.7791 & $-$                          \\
$f_{BCG}/f_{tot}$            & -0.1852          & $-$               & 0.1829  & $-$                          \\
$z$                          & -3.8577           & -0.3301          & $-$      & $-$                          \\
$\mathrm{log_{10}}M_{X,500}$ & $-$               & $-$               & $-$      & $-$                         
\end{tabular}
\label{t:p2logm500}
\end{table}
\section*{Appendix C - Test of Stellar Population on ICL-to-BCG Assignment Dependence}\label{appendixc}
\par It was realised when carrying out the isophotal method of measuring ICL flux that choice of measurement band may potentially be a concern. \cite{2019arXiv191201616M} point out that measuring the ICL in bluer bands can lead to a stronger redshift trend than reality, due to the rapid fade of bluer stellar populations at high redshifts. Although we measure all clusters in the HSC $i$-band, we $k$-correct to the $B$-band, so performed a check on how this effect may influence our results via looking at the trend of the square root of the isophotal areas of the cluster BCGs (i.e. an approximation of the radius beyond which the ICL is considered for our clusters), normalised by the $R_{X,500}$ value of each cluster (to account for cluster size) with redshift. 
\begin{figure}
    \centering
	\includegraphics[width=\columnwidth]{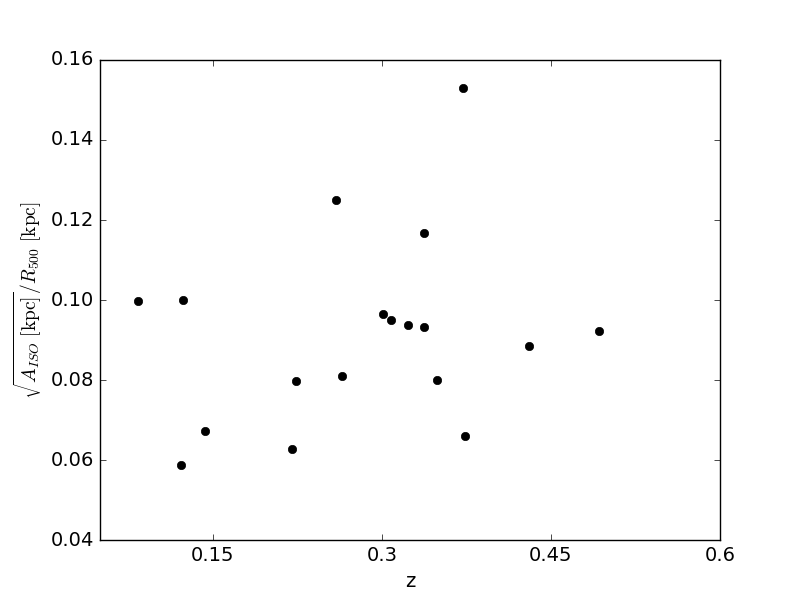}
    \caption{The square root of the isophotal area for each BCG against redshift, normalised by cluster radius. A Spearman rank reveals no strong correlation ($r_{s} = 0.11$, $p = 0.66$).}
    \label{fig:root_area_stelpop}
\end{figure}
\par The result is shown in Figure \ref{fig:root_area_stelpop}. There is a large amount of scatter, and the corresponding Spearman rank does not show evidence for a significant correlation here ($r_{s} = 0.11$, $p = 0.66$). Studies with larger sample sizes may help clarify whether isophotal methods at different wavebands must consider this effect more carefully when interpreting their results.

\label{lastpage}
\end{document}